\documentclass[%
 reprint,
superscriptaddress,
 amsmath,amssymb,
 aps,prl]{revtex4-2}

\usepackage{xurl}
\usepackage[hidelinks,breaklinks]{hyperref}

\usepackage{graphicx}

\title{Stabilizing the inverted phase of a WSe$_2$/BLG/WSe$_2$ heterostructure via hydrostatic pressure}

\begin{document}
\author{M\'at\'e Kedves}
\affiliation{Department of Physics, Institute of Physics, Budapest University of Technology and Economics, M\H uegyetem rkp.\ 3., H-1111 Budapest, Hungary}
\affiliation{MTA-BME Correlated van der Waals Structures Momentum Research Group, M\H uegyetem rkp.\ 3., H-1111 Budapest, Hungary}

\author{B\'alint Szentp\'eteri}
\affiliation{Department of Physics, Institute of Physics, Budapest University of Technology and Economics, M\H uegyetem rkp.\ 3., H-1111 Budapest, Hungary}
\affiliation{MTA-BME Correlated van der Waals Structures Momentum Research Group, M\H uegyetem rkp.\ 3., H-1111 Budapest, Hungary}

\author{Albin M\'arffy}
\affiliation{Department of Physics, Institute of Physics, Budapest University of Technology and Economics, M\H uegyetem rkp.\ 3., H-1111 Budapest, Hungary}
\affiliation{MTA-BME Superconducting Nanoelectronics Momentum Research Group, M\H uegyetem rkp.\ 3., H-1111 Budapest, Hungary}

\author{Endre T\'ov\'ari}
\affiliation{Department of Physics, Institute of Physics, Budapest University of Technology and Economics, M\H uegyetem rkp.\ 3., H-1111 Budapest, Hungary}
\affiliation{MTA-BME Correlated van der Waals Structures Momentum Research Group, M\H uegyetem rkp.\ 3., H-1111 Budapest, Hungary}

\author{Nikos Papadopoulos}
\affiliation{QuTech and Kavli Institute of Nanoscience, Delft University of Technology, 2600 GA Delft, The Netherlands}

\author{Prasanna K. Rout}
\affiliation{QuTech and Kavli Institute of Nanoscience, Delft University of Technology, 2600 GA Delft, The Netherlands}

\author{Kenji Watanabe}
\affiliation{Research Center for Functional Materials, National Institute for Materials Science, 1-1 Namiki, Tsukuba 305-0044, Japan}

\author{Takashi Taniguchi}
\affiliation{International Center for Materials Nanoarchitectonics,
National Institute for Materials Science, 1-1 Namiki, Tsukuba 305-0044, Japan}

\author{Srijit Goswami}
\affiliation{QuTech and Kavli Institute of Nanoscience, Delft University of Technology, 2600 GA Delft, The Netherlands}

\author{Szabolcs Csonka}
\affiliation{Department of Physics, Institute of Physics, Budapest University of Technology and Economics, M\H uegyetem rkp.\ 3., H-1111 Budapest, Hungary}
\affiliation{MTA-BME Superconducting Nanoelectronics Momentum Research Group, M\H uegyetem rkp.\ 3., H-1111 Budapest, Hungary}

\author{P\'eter Makk}
\email{makk.peter@ttk.bme.hu}
\affiliation{Department of Physics, Institute of Physics, Budapest University of Technology and Economics, M\H uegyetem rkp.\ 3., H-1111 Budapest, Hungary}
\affiliation{MTA-BME Correlated van der Waals Structures Momentum Research Group, M\H uegyetem rkp.\ 3., H-1111 Budapest, Hungary}

\begin{abstract}
Bilayer graphene (BLG) was recently shown to host a band-inverted phase with unconventional topology emerging from the Ising-type spin--orbit interaction (SOI) induced by the proximity of transition metal dichalcogenides with large intrinsic SOI. Here, we report the stabilization of this band-inverted phase in BLG symmetrically encapsulated in tungsten-diselenide (WSe$_2$) via hydrostatic pressure. Our observations from low temperature transport measurements are consistent with a single particle model with induced Ising SOI of opposite sign on the two graphene layers. To confirm the strengthening of the inverted phase, we present thermal activation measurements and show that the SOI-induced band gap increases by more than 100\% due to the applied pressure. Finally, the investigation of Landau level spectra reveals the dependence of the level-crossings on the applied magnetic field, which further confirms the enhancement of SOI with pressure.
\end{abstract}

\maketitle

Van der Waals (VdW) engineering provides a powerful method to realize electronic devices with novel functionalities via the combination of multiple 2D materials\,\cite{Geim2013}. An exciting example is the case of graphene connected to materials with large intrinsic spin--orbit interaction (SOI), which allows the generation of an enhanced SOI in graphene via proximity effect\,\cite{Konschuh2012,Gmitra2015,Khoo2017,Garcia2018,Li2019,Zollner2021,Naimer2021,Herling2020,Sierra2021,Peterfalvi2022,Avsar2014,Wang2015,Wang2016,Ghiasi2017,Voelkl2017,Benitez2017,Zihlmann2018,Wakamura2018,Ghiasi2019,Island2019,Wang2019,Omar2019,Tiwari2021,InglaAynes2021,Amann2022}. This, on the one hand, is compelling in the case of spintronics devices since the large spin diffusion length in graphene heterostructures\,\cite{InglaAynes2015,Droegeler2016,Singh2016} could be complemented with electrical tunability\,\cite{Yang2016,Dankert2017,Omar2018} or charge-to-spin conversion effects\,\cite{Garcia2017}. Moreover, it is also interesting from a fundamental point of view since graphene with intrinsic SOI was predicted to be a topological insulator\,\cite{Kane2005}. The observation of increased SOI was demonstrated in the last few years in both single layer\,\cite{Avsar2014,Wang2015,Wang2016,Ghiasi2017,Voelkl2017,Benitez2017,Zihlmann2018,Wakamura2018,Ghiasi2019} and recently in bilayer graphene (BLG)\,\cite{Wang2016,Island2019,Wang2019,Omar2019,Tiwari2021,InglaAynes2021,Amann2022}. It was found that one of the dominating spin--orbit terms is the Ising-type valley-Zeeman term which is an effective magnetic field acting oppositely in the two valleys, and could enable such exciting applications  as a valley-spin valve in BLG\,\cite{Gmitra2017}. Recent compressibility measurements\,\cite{Island2019} have shown that BLG encapsulated in tungsten-diselenide (WSe$_2$) from both sides hosts a band-inverted phase if the sign of induced SOI is different for the two WSe$_2$ layers. In practice, this can be achieved if the twist angle between the two WSe$_2$ layers is e.g. 180$^\circ$\,\cite{David2019,Zollner2021,Peterfalvi2022}.

In this paper, we experimentally investigate the SOI induced in BLG symmetrically encapsulated in WSe$_2$ (WSe$_2$/BLG/WSe$_2$) via transport measurements. We present resistance measurements as a function of charge carrier density ($n$) and transverse displacement field ($D$) at ambient pressure and demonstrate the appearance of the inverted phase (IP). In order to stabilize this phase, we employ our recently developed setup\,\cite{Fueloep2021,Szentpeteri2021} to apply a hydrostatic pressure ($p$) which allows us to decrease the distance between the WSe$_2$ layers and bilayer graphene and to boost the SOI as we have recently demonstrated on single layer graphene\,\cite{Fueloep2021a}. The sample is placed in a piston-cylinder pressure cell where kerosene acts as the pressure mediating medium. More details about this can also be found in Methods. To confirm the increased SOI, we present thermal activation measurements where the evolution of the SOI-induced band gap can be estimated as a function of $D$ and $p$. Finally, we further investigate the induced SOI with quantum Hall measurements by tracking the Landau level crossings as a function of magnetic field.

\begin{figure}
\includegraphics[width=1\columnwidth]{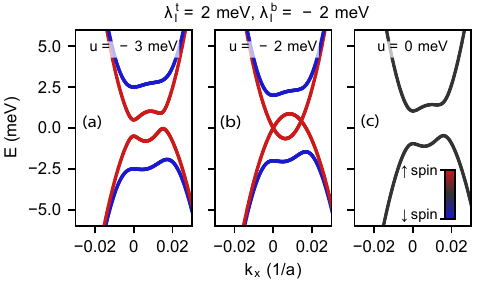}
\caption{\label{fig:bandstructure} a-c) Calculated band structure around the \textbf{K}-point for different values of the interlayer potential difference $u$. Color scale corresponds to the spin polarization of the bands.}
\end{figure}

To reveal the band-inverted phase arising from the Ising SOI in BLG, we show the low-energy band structure of WSe$_2$/BLG/WSe$_2$ in Fig.\,\ref{fig:bandstructure}, calculated using a continuum model by following in the footsteps of Ref.\,\cite{Zollner2021}. The effect of the WSe$_2$ layers in proximity of BLG can be described by the Ising SOI terms $\lambda_{I}^{t}$ and $\lambda_{I}^{b}$ that couple only to the top or bottom layer of BLG and act as a valley-dependent effective magnetic field. For WSe$_2$ layers rotated with respect to each other with 180$^\circ$, the induced SOI couplings will have opposite sign\,\cite{David2019,Zollner2021,Peterfalvi2022}. This is taken into account by the opposite sign of $\lambda_{I}^{t}$ and $\lambda_{I}^{b}$. The transverse displacement field ($D$) in our measurements can be modelled by introducing an interlayer potential difference $u=\frac{-ed}{\epsilon_0\epsilon_{BLG}}D$, where $e$ is the elementary charge, $\epsilon_0$ is the vacuum permittivity, $d=3.3$\,\AA\ is the separation of BLG layers and $\epsilon_{BLG}$ is the effective out-of-plane dielectric constant of BLG.

Fig.\,\ref{fig:bandstructure}.a-c show the calculated band structure around the \textbf{K}-point for different values of $u$, using the parameter values $\lambda_{I}^{t}=-\lambda_{I}^{b}=2$\,meV. Details of the modeling can be found in the Supporting Information. First of all, for $|u|>|\lambda_{I}^{t}|=|\lambda_{I}^{b}|$, we can see the opening of a band gap (Fig.\,\ref{fig:bandstructure}.a), as expected for BLG in a transverse displacement field\,\cite{McCann2006a,Castro2007}. On the other hand, as opposed to pristine BLG, the bands are spin-split and the direction of this spin splitting is opposite for the valence and conduction bands. This is a direct consequence of the opposite sign of $\lambda_{I}^{t}$ and $\lambda_{I}^{b}$ as the valence and conduction bands are localised on different layers due to the large $u$. The band structure in the \textbf{K'}-valley is similar, except that the spin-splitting is reversed due to time reversal symmetry. For $\left|u\right|=|\lambda_{I}^{t,b}|$ (Fig.\,\ref{fig:bandstructure}.b), the $u$-induced band gap approximately equals the spin splitting induced by the Ising SOI and the bands touch. Finally, for $|u|<|\lambda_{I}^{t,b}|$ (Fig.\,\ref{fig:bandstructure}.c), a band gap re-opens and we observe spin-degenerate bands for $u=0$, separated by a gap comparable in size to the Ising SOI terms $\left(\Delta \approx \left|\lambda_{I}^{t}-\lambda_{I}^{b}\right|/2\right)$. This gapped phase is distinct from the band insulating phase at large $u$ in that the valence and conduction bands are no longer layer polarised, hence it is usually referred to as inverted phase (IP). It is worth mentioning that the IP at $|u|<\left|\lambda_{I}^{t}\right|$ is weakly topological unlike the trivial band insulating phase\,\cite{Penaranda2023,Zaletel2019}.

\begin{figure}
\includegraphics[width=1\columnwidth]{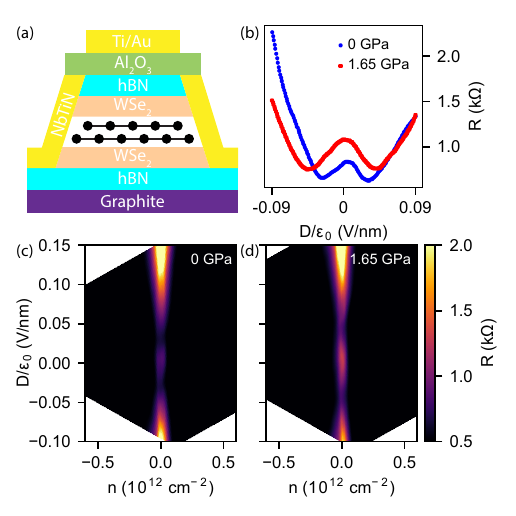}
\caption{\label{fig:resistance} a) Schematic representation of the measured device. Bilayer graphene is symmetrically encapsulated in WSe$_2$ and hBN. b) Line trace of the four-terminal resistance along the CNL for ambient pressure (blue) and $p=1.65$\,GPa (red). c,d) Four-terminal resistance map as a function of charge carrier density $n$ and displacement field $D$ measured at c) ambient pressure and d) an applied pressure of 1.65 GPa. The alternating low and high resistance regions along the CNL indicate the closing and re-opening of a band gap in the bilayer graphene.}
\end{figure}

Our device consists of a BLG flake encapsulated in WSe$_2$ and hexagonal boron nitride (hBN) on both sides, as it is illustrated in Fig.\,\ref{fig:resistance}.a. To enable transport measurements, we fabricated NbTiN edge contacts in a Hall bar geometry. The device also features a graphite bottomgate and a metallic topgate that allow the independent tuning of $n$ and $D$. See Supporting Information for more details about sample fabrication and geometry. The results on similar devices with very similar findings are also shown in the Supporting Information.

Fig.\,\ref{fig:resistance}.c shows the resistance measured in a four-terminal geometry as a function of $n$ and $D$ at ambient pressure at 1.4 K temperature. As expected for BLG, we observe the opening of a band gap at large displacement fields along the charge neutrality line (CNL) at $n=0$, indicated by an increase of resistance. In accordance with the theoretical model and previous compressibility measurements\,\cite{Island2019}, we also observe two local minima separated by a resistance peak at $D=0$ in agreement with the closing and re-opening of the band gap signalling the transition between the band insulator and the IP. This observation is further emphasized in Fig.\,\ref{fig:resistance}.b, where a line trace (blue) of the resistance is shown as a function of $D$, measured along the CNL. It is important to note that during the fabrication process the rotation of WSe$_2$ layers was not controlled. However, from theoretical predictions\,\cite{David2019,Zollner2021,Peterfalvi2022}, we only expect to observe signatures of the IP for a suitable range of rotation angles between the two WSe$_2$ layers (e.g. $\sim$180$^{\circ}$). This is further supported by the fact that not all devices fabricated showed the IP. An example for this case is shown in the Supporting Information where only the band insulating regime can be observed in the resistance map.

To boost the induced SOI and stabilize the IP, we applied a hydrostatic pressure of $p=1.65$ GPa and repeated the previous measurement. Fig.\,\ref{fig:resistance}.d shows the $n$--$D$ map of the resistance after applying the pressure. Although the basic features of the resistance map are similar, two consequences of applying the pressure are clearly visible. First, as it is also illustrated in Fig.\,\ref{fig:resistance}.b, the peak resistance in the IP at $D=0$ increased by $\sim$25\%. Secondly, the displacement field required to close the gap of the IP increased significantly, by about 70\%. Both of these observations can be accounted for by an increase of the Ising SOI term that results in a larger gap at $D=0$ and subsequently in a larger displacement field needed to close the gap. Altough the shift of resistance minima could be explained by the increase of $\epsilon_{BLG}$ or the decrease of interlayer separation $d$, these altogether are not expected to have greater effect than $\sim$20\% \cite{Yankowitz2018,Carr2018}. It is also worth mentioning that the lever arms also change due to the applied pressure, changing the conversion from gate voltages to $n$ and $D$, however, we have corrected for this effect by experimentally determining them from quantum Hall measurements (see Supporting Information).

\begin{figure}
\includegraphics[width=1\columnwidth]{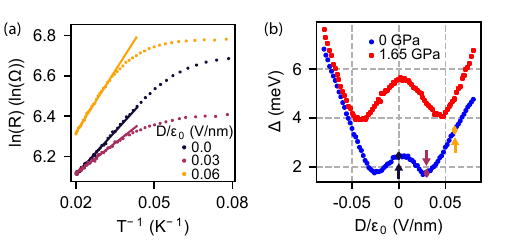}
\caption{\label{fig:activation} Thermal activation measurements along the charge neutrality line. a) Arrhenius plot of the resistance at ambient pressure for selected values of $D$. Solid lines are fits to the linear parts of the data from which the band gap values were obtained. b) Gap $\Delta$ as a function of displacement field at ambient pressure (blue) and an applied pressure of 1.65 GPa. Arrows indicate the $D$ values for which the activation data is shown in a).}
\end{figure}

To quantify the increase of SOI gap due to hydrostatic pressure, we performed thermal activation measurements along the CNL for several values of $D$. Fig.\,\ref{fig:activation}.a demonstrates the evolution of resistance as a function of 1/$T$ for selected values of $D$ at ambient pressure. From this, we extract the band gap using a fit to the high-temperature, linear part of the data where thermal activation -- $\ln(R)\propto \Delta/2k_{B}T$ -- dominates over hopping-related effects\,\cite{Sui2015}. Fig.\,\ref{fig:activation}.b shows the extracted gap values as a function of $D$ with and without applied pressure. First of all, a factor of 2 increase is clearly visible in the gap at $D=0$ for $p=1.65$\,GPa, that is consistent with the observed increase of resistance. Secondly, the higher $D$ needed to reach the gap minima is also confirmed. We also note that the band gap cannot be fully closed which we attribute to spatial inhomogeneity in the sample.

\begin{figure*}[!ht]
\includegraphics[width=2\columnwidth]{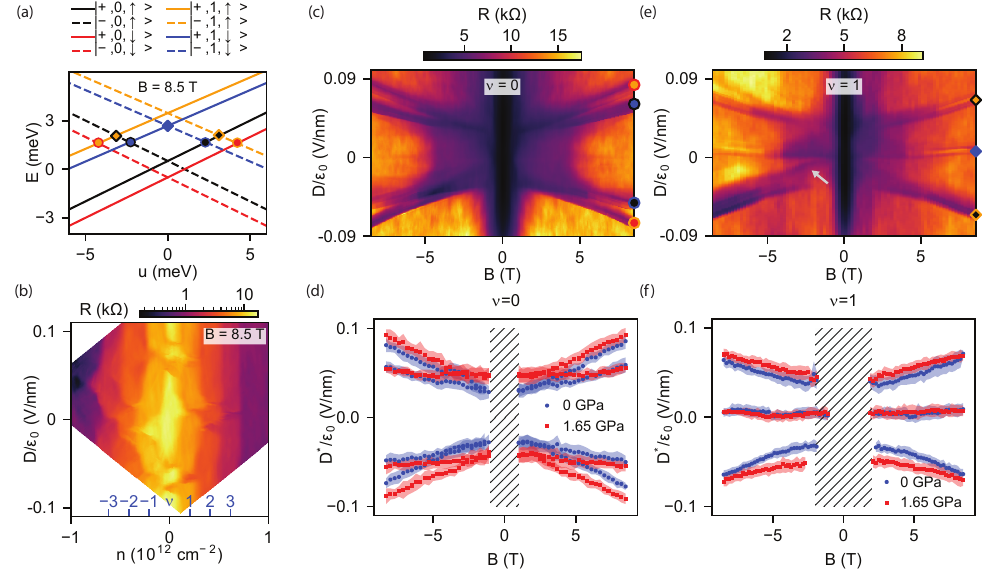}
\caption{\label{fig:QHall}a) Low energy Landau level spectrum at $B=8.5$ T obtained from single-particle continuum model with $\lambda_{I}^{t}=-\lambda_{I}^{b}=2$\,meV. b) Four-terminal resistance as a function of $n$ and $D$ measured at $B=8.5$\,T out-of-plane magnetic field and ambient pressure. Resistance plateaus correspond to different $\nu$ filling factors. Abrupt changes in resistance at a given $\nu$ as a function of $D$ indicate the crossings of LLs. c,e) Measurements of LL crossings as a function of $B$ for $\nu=0$ and $\nu=1$, respectively, for $p=0$. Symbols denote LL crossings shown in (a). d,f) Critical displacement field $D^{*}$ corresponding to LL crossings for $\nu=0$ and $\nu=1$ extracted from $D-B$ maps measured at $p=0$ (blue, see (c,e)) and $p=1.65$ GPa (red).}
\end{figure*}

The experimentally determined band gaps allow us to quantify the SOI parameters. By adjusting the theoretical model to match the positions of the gap minima and the opening of the trivial gap for $p=0$, we extract $\lambda_{I}^{t}=-\lambda_{I}^{b}=2.2\pm0.4$\,meV. Similarly, we can extract the SOI parameters at $p=1.65$\,GPa. For these, we obtain $\lambda_{I}^{t}=-\lambda_{I}^{b}=5.6\pm0.6$\,meV. The SOI parameters extracted from the minima give the same order of magnitude estimate as the gaps at $D=0$ extracted from thermal activation directly. A more detailed discussion on the extraction and possible errors is given in the Supporting Information. We expect that all layer distances (e.g. hBN-hBN, BLG-WSe$_2$ and $d$) change due to the applied pressure as it is also reflected in the change of lever arms. The extracted increase of SOI strength due to the change of BLG-WSe$_2$ distances is consistent with theoretical predictions in Ref.\,\cite{Fueloep2021} where almost a factor of 3 increase was predicted for an applied pressure of 1.8\,GPa. Importantly, we have found similar results in two further devices shown in the Supporting Information.

The quantum Hall effect in BLG provides us another tool to investigate the Ising SOI induced by the WSe$_2$ layers. The two-fold degeneracy of valley isospin ($\xi=+,-$), the first two orbitals ($N=0,1$) and spin ($\sigma=\uparrow,\downarrow$) give rise to an eight-fold degenerate Landau level (LL) near zero-energy\,\cite{McCann2006,Novoselov2006,Hunt2017}. This degeneracy is weakly lifted by the interlayer potential difference, Zeeman energy, coupling elements between the BLG layers\,\cite{Khoo2018} and the induced Ising SOI\,\cite{Wang2019}. We can obtain the energy spectrum of this set of eight closely-spaced sublevels -- labeled by $|\xi ,N,\sigma \rangle$ -- by introducing a perpendicular magnetic field in our continuum model, as detailed in \cite{Khoo2018}. This is shown in Fig.\,\ref{fig:QHall}.a for $B=8.5$\,T as a function of the interlayer potential ($u$). LLs with different $\xi$ reside on different layers of the BLG, therefore $u$ induces a splitting between these levels. Secondly, the finite magnetic field causes the Zeeman-splitting of levels with different $\sigma$. Finally, the Ising SOI induces an additional effective Zeeman field associated to a given layer, further splitting the levels. The key feature that should be noted here, is that for a given filling factor $\nu$, crossings of LLs can be observed and the position of crossing points along the $u$ axis depend on SOI parameters as well as on the magnetic field. These level crossings manifest as sudden changes of resistance in our transport measurements as it is illustrated in Fig.\,\ref{fig:QHall}.b. Here, the $n-D$ map of the resistance is shown as measured at $B=8.5$\,T with fully developed resistance plateaus (due to the unconvetional geometry, see Supporting Information) corresponding to the sublevels of $\nu \in \left[ -4,4\right]$. For a given filling factor $\nu$, we observe $4-|\nu|$ different $D$ values where the resistance deviates from the surrounding plateau corresponding to the crossing of LLs, as expected from the model. 

The evolution of LL crossings with $B$ can be observed by performing measurements at fixed filling factors, as it is shown in Fig.\,\ref{fig:QHall}.c and \ref{fig:QHall}.e for $\nu=0$ and $\nu=1$, respectively. During the latter measurement, the carrier density $n$ was tuned such that the filling factor given by $\nu=nh/eB$ was kept constant. On both panels, we can observe $4-\nu$ LL crossings that evolve as $B$ is tuned, until they disappear at low magnetic fields where we can no longer resolve LL plateaus. This $B$-dependent behaviour enables us to investigate the effect of SOI on the LL structure. Fig.\,\ref{fig:QHall}.d and \ref{fig:QHall}.f shows the critical displacement field $D^{*}$ values -- where LL crossings can be observed -- extracted from Fig.\,\ref{fig:QHall}.c and \ref{fig:QHall}.e and similar maps measured at $p=1.65$\,GPa (see Supporting Information). For $\nu=0$ (Fig.\,\ref{fig:QHall}.d), the most important observation is that the crossing points do not extrapolate to zero as $B\rightarrow 0$\,T which is a direct consequence of the induced Ising SOI. It is also clearly visible that due to the applied pressure, $|D^{*}|$ is generally increased, especially at lower $B$-fields, indicating that the Ising SOI has increased, in agreement with our thermal activation measurements. For $\nu=1$ (Fig.\,\ref{fig:QHall}.f), similar trends can be observed. The two LL crossings at finite $D$ saturate for small $B$, while the third crossing remains at $D=0$. We note that the $D^{*}(B)$ curves for $p=1.65$ GPa cannot be scaled down to the $p=0$ curves, which confirms that our observations cannot simply be explained by an increased $\epsilon_{BLG}$ or decreased interlayer separation distance, but are the results of enhanced SOI. We also point out that some lines which extrapolate to $D=0$ can also be observed (e.g. Fig.\,\ref{fig:QHall}.e, grey arrow). This could also be explained by sample inhomogeneity. It is also important to note that our single-particle model fails to quantitatively predict the $B$-dependence of the LL crossings indicating the importance of electron-electron interactions (see Supporting Information). 

In conclusion, we showed that the IP observed in BLG symmetrically encapsulated between twisted WSe$_2$ layers can be stabilized by applying hydrostatic pressure which enhances the proximity induced SOI. We presented thermal activation measurements as a means to quantify the Ising SOI parameters in this system and showed an increase of approximately 150\% due to the applied pressure. In order to gain more information on the twist angle dependence of the SOI, a more systematic study with several samples with well-controlled twist angles is needed. The enhancement of Ising SOI with pressure was further confirmed from quantum Hall measurements. However, to extract SOI strengths from these measurements, more complex models are needed that also take into account interaction effects. Our study shows that the hydrostatic pressure is an efficient tuning knob to control the induced Ising SOI, thereby the topological phase in WSe$_2$/BLG/WSe$_2$. 

The IP has a distinct topology from the band insulator phase at large $D$, and the presence of edge states are expected\,\cite{Penaranda2023}. The presence of these states should be studied in better defined sample geometries\,\cite{SanchezYamagishi2016,Veyrat2020} or using supercurrent interferometry\,\cite{Hart2014,Indolese2018}. Opposed to the weak protection of the edge states in this system, a strong topological insulator phase is predicted in ABC trilayer graphene\,\cite{Li2012,Zaletel2019}. Furthermore, pressure could also be used in case of magic-angle twisted BLG, in which topological phase transitions between different Chern insulator states are expected as a function of SOI strength\,\cite{Wang2020}.

\section*{Methods}
\subsection{Sample fabrication}
The dry-transfer technique with PC/PDMS hemispheres is employed to stack hBN (35 nm)/WSe$_{2}$ (19 nm)/BLG/WSe$_{2}$ (19 nm)/hBN (60 nm)/graphite. To fabricate electrical contacts to the Hall bar, we use e-beam lithography patterning followed by a reactive ion etching step using CHF3/O2 mixture and finally deposit Ti (5nm)/NbTiN (100 nm) by dc sputtering. We deposit Al2O3 (30 nm) using ALD which acts as the gate dielectric and isolates the ohmic contacts from the top gate. Finally, the top gate is defined by e-beam lithography and deposition of Ti (5 nm)/Au (100 nm).

\subsection{Transport measurements}
Transport measurements were carried out in an Oxford cryostat equipped with a variable temperature insert (VTI) at a base temperature of 1.4 K (unless otherwise stated). Measurements were performed using lock-in technique at 1.17\,kHz frequency.

\subsection{Pressurization}
The sample is first bonded to a high pressure sample holder and placed in a piston-cylinder pressure cell where kerosene acts
as the pressure mediating medium. To change the applied pressure the sample is warmed up to room temperature where the pressure is applied using a hydraulic press and the sample
is cooled down again. Our pressure cell is described in more detail in Ref.\,\cite{Fueloep2021a}

\subsection{Data availability}
Source data of the measurements and the Python code for the simulation are publicly available at https://doi.org/10.5281/zenodo.8406628.

\section*{Author contributions}
N.P. and P.K.R fabricated the device. Measurements were performed by M.K., B.Sz., P.K.R. with the help of M.A., P.M. M.K. and B.Sz. did the data analysis. B.Sz. did the theoretical calculation. M.K., B.Sz. and E.T. and P.M. wrote the paper and all authors discussed the results and worked on the manuscript. K.W. and T.T. grew the hBN crystals. The project was guided by Sz.Cs., S.G. and P.M.

\section*{acknowledgement}
This work acknowledges support from the Topograph, MultiSpin and 2DSOTECH FlagERA networks, the OTKA K138433 and PD 134758 grants and the VEKOP 2.3.3-15-2017-00015 grant. This research was supported by the Ministry of Culture and Innovation and the National Research, Development and Innovation Office within the Quantum Information National Laboratory of Hungary (Grant No. 2022-2.1.1-NL-2022-00004), by SuperTop QuantERA network, by the FET Open AndQC netwrok. We acknowledge COST Action CA 21144 superQUMAP. P.M. and E.T. received funding from Bolyai Fellowship. This project was supported by the ÚNKP-22-3-II New National Excellence Program of the Ministry for Innovation and Technology from the source of the National Research, Development and Innovation Found. K.W. and T.T. acknowledge support from JSPS KAKENHI (Grant Numbers 19H05790, 20H00354 and 21H05233). The authors thank Pablo San-Jose, Elsa Prada and Fernando Pe\~{n}aranda for fruitful discussions.

\bibliography{inverted_gap}

\begin{thebibliography}{56}%
\makeatletter
\providecommand \@ifxundefined [1]{%
 \@ifx{#1\undefined}
}%
\providecommand \@ifnum [1]{%
 \ifnum #1\expandafter \@firstoftwo
 \else \expandafter \@secondoftwo
 \fi
}%
\providecommand \@ifx [1]{%
 \ifx #1\expandafter \@firstoftwo
 \else \expandafter \@secondoftwo
 \fi
}%
\providecommand \natexlab [1]{#1}%
\providecommand \enquote  [1]{``#1''}%
\providecommand \bibnamefont  [1]{#1}%
\providecommand \bibfnamefont [1]{#1}%
\providecommand \citenamefont [1]{#1}%
\providecommand \href@noop [0]{\@secondoftwo}%
\providecommand \href [0]{\begingroup \@sanitize@url \@href}%
\providecommand \@href[1]{\@@startlink{#1}\@@href}%
\providecommand \@@href[1]{\endgroup#1\@@endlink}%
\providecommand \@sanitize@url [0]{\catcode `\\12\catcode `\$12\catcode `\&12\catcode `\#12\catcode `\^12\catcode `\_12\catcode `\%12\relax}%
\providecommand \@@startlink[1]{}%
\providecommand \@@endlink[0]{}%
\providecommand \url  [0]{\begingroup\@sanitize@url \@url }%
\providecommand \@url [1]{\endgroup\@href {#1}{\urlprefix }}%
\providecommand \urlprefix  [0]{URL }%
\providecommand \Eprint [0]{\href }%
\providecommand \doibase [0]{https://doi.org/}%
\providecommand \selectlanguage [0]{\@gobble}%
\providecommand \bibinfo  [0]{\@secondoftwo}%
\providecommand \bibfield  [0]{\@secondoftwo}%
\providecommand \translation [1]{[#1]}%
\providecommand \BibitemOpen [0]{}%
\providecommand \bibitemStop [0]{}%
\providecommand \bibitemNoStop [0]{.\EOS\space}%
\providecommand \EOS [0]{\spacefactor3000\relax}%
\providecommand \BibitemShut  [1]{\csname bibitem#1\endcsname}%
\let\auto@bib@innerbib\@empty
%</preamble>
\bibitem [{\citenamefont {Geim}\ and\ \citenamefont {Grigorieva}(2013)}]{Geim2013}%
  \BibitemOpen
  \bibfield  {author} {\bibinfo {author} {\bibfnamefont {A.~K.}\ \bibnamefont {Geim}}\ and\ \bibinfo {author} {\bibfnamefont {I.~V.}\ \bibnamefont {Grigorieva}},\ }\bibfield  {title} {\bibinfo {title} {Van der waals heterostructures},\ }\href {https://doi.org/10.1038/nature12385} {\bibfield  {journal} {\bibinfo  {journal} {Nature}\ }\textbf {\bibinfo {volume} {499}},\ \bibinfo {pages} {419} (\bibinfo {year} {2013})}\BibitemShut {NoStop}%
\bibitem [{\citenamefont {Konschuh}\ \emph {et~al.}(2012)\citenamefont {Konschuh}, \citenamefont {Gmitra}, \citenamefont {Kochan},\ and\ \citenamefont {Fabian}}]{Konschuh2012}%
  \BibitemOpen
  \bibfield  {author} {\bibinfo {author} {\bibfnamefont {S.}~\bibnamefont {Konschuh}}, \bibinfo {author} {\bibfnamefont {M.}~\bibnamefont {Gmitra}}, \bibinfo {author} {\bibfnamefont {D.}~\bibnamefont {Kochan}},\ and\ \bibinfo {author} {\bibfnamefont {J.}~\bibnamefont {Fabian}},\ }\bibfield  {title} {\bibinfo {title} {Theory of spin-orbit coupling in bilayer graphene},\ }\href {https://doi.org/10.1103/PhysRevB.85.115423} {\bibfield  {journal} {\bibinfo  {journal} {Phys. Rev. B}\ }\textbf {\bibinfo {volume} {85}},\ \bibinfo {pages} {115423} (\bibinfo {year} {2012})}\BibitemShut {NoStop}%
\bibitem [{\citenamefont {Gmitra}\ and\ \citenamefont {Fabian}(2015)}]{Gmitra2015}%
  \BibitemOpen
  \bibfield  {author} {\bibinfo {author} {\bibfnamefont {M.}~\bibnamefont {Gmitra}}\ and\ \bibinfo {author} {\bibfnamefont {J.}~\bibnamefont {Fabian}},\ }\bibfield  {title} {\bibinfo {title} {Graphene on transition-metal dichalcogenides: A platform for proximity spin-orbit physics and optospintronics},\ }\href {https://doi.org/10.1103/PhysRevB.92.155403} {\bibfield  {journal} {\bibinfo  {journal} {Phys. Rev. B}\ }\textbf {\bibinfo {volume} {92}},\ \bibinfo {pages} {155403} (\bibinfo {year} {2015})}\BibitemShut {NoStop}%
\bibitem [{\citenamefont {Khoo}\ \emph {et~al.}(2017)\citenamefont {Khoo}, \citenamefont {Morpurgo},\ and\ \citenamefont {Levitov}}]{Khoo2017}%
  \BibitemOpen
  \bibfield  {author} {\bibinfo {author} {\bibfnamefont {J.~Y.}\ \bibnamefont {Khoo}}, \bibinfo {author} {\bibfnamefont {A.~F.}\ \bibnamefont {Morpurgo}},\ and\ \bibinfo {author} {\bibfnamefont {L.}~\bibnamefont {Levitov}},\ }\bibfield  {title} {\bibinfo {title} {On-demand spin{\textendash}orbit interaction from which-layer tunability in bilayer graphene},\ }\href {https://doi.org/10.1021/acs.nanolett.7b03604} {\bibfield  {journal} {\bibinfo  {journal} {Nano Letters}\ }\textbf {\bibinfo {volume} {17}},\ \bibinfo {pages} {7003} (\bibinfo {year} {2017})}\BibitemShut {NoStop}%
\bibitem [{\citenamefont {Garcia}\ \emph {et~al.}(2018)\citenamefont {Garcia}, \citenamefont {Vila}, \citenamefont {Cummings},\ and\ \citenamefont {Roche}}]{Garcia2018}%
  \BibitemOpen
  \bibfield  {author} {\bibinfo {author} {\bibfnamefont {J.~H.}\ \bibnamefont {Garcia}}, \bibinfo {author} {\bibfnamefont {M.}~\bibnamefont {Vila}}, \bibinfo {author} {\bibfnamefont {A.~W.}\ \bibnamefont {Cummings}},\ and\ \bibinfo {author} {\bibfnamefont {S.}~\bibnamefont {Roche}},\ }\bibfield  {title} {\bibinfo {title} {Spin transport in graphene/transition metal dichalcogenide heterostructures},\ }\href {https://doi.org/10.1039/c7cs00864c} {\bibfield  {journal} {\bibinfo  {journal} {Chemical Society Reviews}\ }\textbf {\bibinfo {volume} {47}},\ \bibinfo {pages} {3359} (\bibinfo {year} {2018})}\BibitemShut {NoStop}%
\bibitem [{\citenamefont {Li}\ and\ \citenamefont {Koshino}(2019)}]{Li2019}%
  \BibitemOpen
  \bibfield  {author} {\bibinfo {author} {\bibfnamefont {Y.}~\bibnamefont {Li}}\ and\ \bibinfo {author} {\bibfnamefont {M.}~\bibnamefont {Koshino}},\ }\bibfield  {title} {\bibinfo {title} {Twist-angle dependence of the proximity spin-orbit coupling in graphene on transition-metal dichalcogenides},\ }\href {https://doi.org/10.1103/PhysRevB.99.075438} {\bibfield  {journal} {\bibinfo  {journal} {Phys. Rev. B}\ }\textbf {\bibinfo {volume} {99}},\ \bibinfo {pages} {075438} (\bibinfo {year} {2019})}\BibitemShut {NoStop}%
\bibitem [{\citenamefont {Zollner}\ and\ \citenamefont {Fabian}(2021)}]{Zollner2021}%
  \BibitemOpen
  \bibfield  {author} {\bibinfo {author} {\bibfnamefont {K.}~\bibnamefont {Zollner}}\ and\ \bibinfo {author} {\bibfnamefont {J.}~\bibnamefont {Fabian}},\ }\bibfield  {title} {\bibinfo {title} {Bilayer graphene encapsulated within monolayers of ${\mathrm{ws}}_{2}$ or ${\mathrm{cr}}_{2}{\mathrm{ge}}_{2}{\mathrm{te}}_{6}$: Tunable proximity spin-orbit or exchange coupling},\ }\href {https://doi.org/10.1103/PhysRevB.104.075126} {\bibfield  {journal} {\bibinfo  {journal} {Phys. Rev. B}\ }\textbf {\bibinfo {volume} {104}},\ \bibinfo {pages} {075126} (\bibinfo {year} {2021})}\BibitemShut {NoStop}%
\bibitem [{\citenamefont {Naimer}\ \emph {et~al.}(2021)\citenamefont {Naimer}, \citenamefont {Zollner}, \citenamefont {Gmitra},\ and\ \citenamefont {Fabian}}]{Naimer2021}%
  \BibitemOpen
  \bibfield  {author} {\bibinfo {author} {\bibfnamefont {T.}~\bibnamefont {Naimer}}, \bibinfo {author} {\bibfnamefont {K.}~\bibnamefont {Zollner}}, \bibinfo {author} {\bibfnamefont {M.}~\bibnamefont {Gmitra}},\ and\ \bibinfo {author} {\bibfnamefont {J.}~\bibnamefont {Fabian}},\ }\bibfield  {title} {\bibinfo {title} {Twist-angle dependent proximity induced spin-orbit coupling in graphene/transition metal dichalcogenide heterostructures},\ }\href {https://doi.org/10.1103/PhysRevB.104.195156} {\bibfield  {journal} {\bibinfo  {journal} {Phys. Rev. B}\ }\textbf {\bibinfo {volume} {104}},\ \bibinfo {pages} {195156} (\bibinfo {year} {2021})}\BibitemShut {NoStop}%
\bibitem [{\citenamefont {Herling}\ \emph {et~al.}(2020)\citenamefont {Herling}, \citenamefont {Safeer}, \citenamefont {Ingla-Aynés}, \citenamefont {Ontoso}, \citenamefont {Hueso},\ and\ \citenamefont {Casanova}}]{Herling2020}%
  \BibitemOpen
  \bibfield  {author} {\bibinfo {author} {\bibfnamefont {F.}~\bibnamefont {Herling}}, \bibinfo {author} {\bibfnamefont {C.~K.}\ \bibnamefont {Safeer}}, \bibinfo {author} {\bibfnamefont {J.}~\bibnamefont {Ingla-Aynés}}, \bibinfo {author} {\bibfnamefont {N.}~\bibnamefont {Ontoso}}, \bibinfo {author} {\bibfnamefont {L.~E.}\ \bibnamefont {Hueso}},\ and\ \bibinfo {author} {\bibfnamefont {F.}~\bibnamefont {Casanova}},\ }\bibfield  {title} {\bibinfo {title} {Gate tunability of highly efficient spin-to-charge conversion by spin hall effect in graphene proximitized with wse$_{2}$},\ }\href {https://doi.org/10.1063/5.0006101} {\bibfield  {journal} {\bibinfo  {journal} {APL Materials}\ }\textbf {\bibinfo {volume} {8}},\ \bibinfo {pages} {071103} (\bibinfo {year} {2020})},\ \Eprint {https://arxiv.org/abs/https://doi.org/10.1063/5.0006101} {https://doi.org/10.1063/5.0006101} \BibitemShut {NoStop}%
\bibitem [{\citenamefont {Sierra}\ \emph {et~al.}(2021)\citenamefont {Sierra}, \citenamefont {Fabian}, \citenamefont {Kawakami}, \citenamefont {Roche},\ and\ \citenamefont {Valenzuela}}]{Sierra2021}%
  \BibitemOpen
  \bibfield  {author} {\bibinfo {author} {\bibfnamefont {J.~F.}\ \bibnamefont {Sierra}}, \bibinfo {author} {\bibfnamefont {J.}~\bibnamefont {Fabian}}, \bibinfo {author} {\bibfnamefont {R.~K.}\ \bibnamefont {Kawakami}}, \bibinfo {author} {\bibfnamefont {S.}~\bibnamefont {Roche}},\ and\ \bibinfo {author} {\bibfnamefont {S.~O.}\ \bibnamefont {Valenzuela}},\ }\bibfield  {title} {\bibinfo {title} {Van der waals heterostructures for spintronics and opto-spintronics},\ }\href {https://doi.org/10.1038/s41565-021-00936-x} {\bibfield  {journal} {\bibinfo  {journal} {Nature Nanotechnology}\ }\textbf {\bibinfo {volume} {16}},\ \bibinfo {pages} {856} (\bibinfo {year} {2021})}\BibitemShut {NoStop}%
\bibitem [{\citenamefont {P\'eterfalvi}\ \emph {et~al.}(2022)\citenamefont {P\'eterfalvi}, \citenamefont {David}, \citenamefont {Rakyta}, \citenamefont {Burkard},\ and\ \citenamefont {Korm\'anyos}}]{Peterfalvi2022}%
  \BibitemOpen
  \bibfield  {author} {\bibinfo {author} {\bibfnamefont {C.~G.}\ \bibnamefont {P\'eterfalvi}}, \bibinfo {author} {\bibfnamefont {A.}~\bibnamefont {David}}, \bibinfo {author} {\bibfnamefont {P.}~\bibnamefont {Rakyta}}, \bibinfo {author} {\bibfnamefont {G.}~\bibnamefont {Burkard}},\ and\ \bibinfo {author} {\bibfnamefont {A.}~\bibnamefont {Korm\'anyos}},\ }\bibfield  {title} {\bibinfo {title} {Quantum interference tuning of spin-orbit coupling in twisted van der waals trilayers},\ }\href {https://doi.org/10.1103/PhysRevResearch.4.L022049} {\bibfield  {journal} {\bibinfo  {journal} {Phys. Rev. Res.}\ }\textbf {\bibinfo {volume} {4}},\ \bibinfo {pages} {L022049} (\bibinfo {year} {2022})}\BibitemShut {NoStop}%
\bibitem [{\citenamefont {Avsar}\ \emph {et~al.}(2014)\citenamefont {Avsar}, \citenamefont {Tan}, \citenamefont {Taychatanapat}, \citenamefont {Balakrishnan}, \citenamefont {Koon}, \citenamefont {Yeo}, \citenamefont {Lahiri}, \citenamefont {Carvalho}, \citenamefont {Rodin}, \citenamefont {O’Farrell}, \citenamefont {Eda}, \citenamefont {Castro~Neto},\ and\ \citenamefont {Özyilmaz}}]{Avsar2014}%
  \BibitemOpen
  \bibfield  {author} {\bibinfo {author} {\bibfnamefont {A.}~\bibnamefont {Avsar}}, \bibinfo {author} {\bibfnamefont {J.~Y.}\ \bibnamefont {Tan}}, \bibinfo {author} {\bibfnamefont {T.}~\bibnamefont {Taychatanapat}}, \bibinfo {author} {\bibfnamefont {J.}~\bibnamefont {Balakrishnan}}, \bibinfo {author} {\bibfnamefont {G.~K.~W.}\ \bibnamefont {Koon}}, \bibinfo {author} {\bibfnamefont {Y.}~\bibnamefont {Yeo}}, \bibinfo {author} {\bibfnamefont {J.}~\bibnamefont {Lahiri}}, \bibinfo {author} {\bibfnamefont {A.}~\bibnamefont {Carvalho}}, \bibinfo {author} {\bibfnamefont {A.~S.}\ \bibnamefont {Rodin}}, \bibinfo {author} {\bibfnamefont {E.~C.~T.}\ \bibnamefont {O’Farrell}}, \bibinfo {author} {\bibfnamefont {G.}~\bibnamefont {Eda}}, \bibinfo {author} {\bibfnamefont {A.~H.}\ \bibnamefont {Castro~Neto}},\ and\ \bibinfo {author} {\bibfnamefont {B.}~\bibnamefont {Özyilmaz}},\ }\bibfield  {title} {\bibinfo {title} {Spin{\textendash}orbit proximity effect in graphene},\ }\href {https://doi.org/10.1038/ncomms5875}
  {\bibfield  {journal} {\bibinfo  {journal} {Nature Communications}\ }\textbf {\bibinfo {volume} {5}},\ \bibinfo {pages} {4875} (\bibinfo {year} {2014})}\BibitemShut {NoStop}%
\bibitem [{\citenamefont {Wang}\ \emph {et~al.}(2015)\citenamefont {Wang}, \citenamefont {Ki}, \citenamefont {Chen}, \citenamefont {Berger}, \citenamefont {MacDonald},\ and\ \citenamefont {Morpurgo}}]{Wang2015}%
  \BibitemOpen
  \bibfield  {author} {\bibinfo {author} {\bibfnamefont {Z.}~\bibnamefont {Wang}}, \bibinfo {author} {\bibfnamefont {D.-K.}\ \bibnamefont {Ki}}, \bibinfo {author} {\bibfnamefont {H.}~\bibnamefont {Chen}}, \bibinfo {author} {\bibfnamefont {H.}~\bibnamefont {Berger}}, \bibinfo {author} {\bibfnamefont {A.~H.}\ \bibnamefont {MacDonald}},\ and\ \bibinfo {author} {\bibfnamefont {A.~F.}\ \bibnamefont {Morpurgo}},\ }\bibfield  {title} {\bibinfo {title} {Strong interface-induced spin{\textendash}orbit interaction in graphene on {WS}2},\ }\href {https://doi.org/10.1038/ncomms9339} {\bibfield  {journal} {\bibinfo  {journal} {Nature Communications}\ }\textbf {\bibinfo {volume} {6}},\ \bibinfo {pages} {8339} (\bibinfo {year} {2015})}\BibitemShut {NoStop}%
\bibitem [{\citenamefont {Wang}\ \emph {et~al.}(2016)\citenamefont {Wang}, \citenamefont {Ki}, \citenamefont {Khoo}, \citenamefont {Mauro}, \citenamefont {Berger}, \citenamefont {Levitov},\ and\ \citenamefont {Morpurgo}}]{Wang2016}%
  \BibitemOpen
  \bibfield  {author} {\bibinfo {author} {\bibfnamefont {Z.}~\bibnamefont {Wang}}, \bibinfo {author} {\bibfnamefont {D.-K.}\ \bibnamefont {Ki}}, \bibinfo {author} {\bibfnamefont {J.~Y.}\ \bibnamefont {Khoo}}, \bibinfo {author} {\bibfnamefont {D.}~\bibnamefont {Mauro}}, \bibinfo {author} {\bibfnamefont {H.}~\bibnamefont {Berger}}, \bibinfo {author} {\bibfnamefont {L.~S.}\ \bibnamefont {Levitov}},\ and\ \bibinfo {author} {\bibfnamefont {A.~F.}\ \bibnamefont {Morpurgo}},\ }\bibfield  {title} {\bibinfo {title} {Origin and magnitude of `designer' spin-orbit interaction in graphene on semiconducting transition metal dichalcogenides},\ }\href {https://doi.org/10.1103/PhysRevX.6.041020} {\bibfield  {journal} {\bibinfo  {journal} {Phys. Rev. X}\ }\textbf {\bibinfo {volume} {6}},\ \bibinfo {pages} {041020} (\bibinfo {year} {2016})}\BibitemShut {NoStop}%
\bibitem [{\citenamefont {Ghiasi}\ \emph {et~al.}(2017)\citenamefont {Ghiasi}, \citenamefont {Ingla-Ayn{\'{e}}s}, \citenamefont {Kaverzin},\ and\ \citenamefont {van Wees}}]{Ghiasi2017}%
  \BibitemOpen
  \bibfield  {author} {\bibinfo {author} {\bibfnamefont {T.~S.}\ \bibnamefont {Ghiasi}}, \bibinfo {author} {\bibfnamefont {J.}~\bibnamefont {Ingla-Ayn{\'{e}}s}}, \bibinfo {author} {\bibfnamefont {A.~A.}\ \bibnamefont {Kaverzin}},\ and\ \bibinfo {author} {\bibfnamefont {B.~J.}\ \bibnamefont {van Wees}},\ }\bibfield  {title} {\bibinfo {title} {Large proximity-induced spin lifetime anisotropy in transition-metal dichalcogenide/graphene heterostructures},\ }\href {https://doi.org/10.1021/acs.nanolett.7b03460} {\bibfield  {journal} {\bibinfo  {journal} {Nano Letters}\ }\textbf {\bibinfo {volume} {17}},\ \bibinfo {pages} {7528} (\bibinfo {year} {2017})}\BibitemShut {NoStop}%
\bibitem [{\citenamefont {V\"olkl}\ \emph {et~al.}(2017)\citenamefont {V\"olkl}, \citenamefont {Rockinger}, \citenamefont {Drienovsky}, \citenamefont {Watanabe}, \citenamefont {Taniguchi}, \citenamefont {Weiss},\ and\ \citenamefont {Eroms}}]{Voelkl2017}%
  \BibitemOpen
  \bibfield  {author} {\bibinfo {author} {\bibfnamefont {T.}~\bibnamefont {V\"olkl}}, \bibinfo {author} {\bibfnamefont {T.}~\bibnamefont {Rockinger}}, \bibinfo {author} {\bibfnamefont {M.}~\bibnamefont {Drienovsky}}, \bibinfo {author} {\bibfnamefont {K.}~\bibnamefont {Watanabe}}, \bibinfo {author} {\bibfnamefont {T.}~\bibnamefont {Taniguchi}}, \bibinfo {author} {\bibfnamefont {D.}~\bibnamefont {Weiss}},\ and\ \bibinfo {author} {\bibfnamefont {J.}~\bibnamefont {Eroms}},\ }\bibfield  {title} {\bibinfo {title} {Magnetotransport in heterostructures of transition metal dichalcogenides and graphene},\ }\href {https://doi.org/10.1103/PhysRevB.96.125405} {\bibfield  {journal} {\bibinfo  {journal} {Phys. Rev. B}\ }\textbf {\bibinfo {volume} {96}},\ \bibinfo {pages} {125405} (\bibinfo {year} {2017})}\BibitemShut {NoStop}%
\bibitem [{\citenamefont {Ben{\'{\i}}tez}\ \emph {et~al.}(2017)\citenamefont {Ben{\'{\i}}tez}, \citenamefont {Sierra}, \citenamefont {Torres}, \citenamefont {Arrighi}, \citenamefont {Bonell}, \citenamefont {Costache},\ and\ \citenamefont {Valenzuela}}]{Benitez2017}%
  \BibitemOpen
  \bibfield  {author} {\bibinfo {author} {\bibfnamefont {L.~A.}\ \bibnamefont {Ben{\'{\i}}tez}}, \bibinfo {author} {\bibfnamefont {J.~F.}\ \bibnamefont {Sierra}}, \bibinfo {author} {\bibfnamefont {W.~S.}\ \bibnamefont {Torres}}, \bibinfo {author} {\bibfnamefont {A.}~\bibnamefont {Arrighi}}, \bibinfo {author} {\bibfnamefont {F.}~\bibnamefont {Bonell}}, \bibinfo {author} {\bibfnamefont {M.~V.}\ \bibnamefont {Costache}},\ and\ \bibinfo {author} {\bibfnamefont {S.~O.}\ \bibnamefont {Valenzuela}},\ }\bibfield  {title} {\bibinfo {title} {Strongly anisotropic spin relaxation in graphene{\textendash}transition metal dichalcogenide heterostructures at room temperature},\ }\href {https://doi.org/10.1038/s41567-017-0019-2} {\bibfield  {journal} {\bibinfo  {journal} {Nature Physics}\ }\textbf {\bibinfo {volume} {14}},\ \bibinfo {pages} {303} (\bibinfo {year} {2017})}\BibitemShut {NoStop}%
\bibitem [{\citenamefont {Zihlmann}\ \emph {et~al.}(2018)\citenamefont {Zihlmann}, \citenamefont {Cummings}, \citenamefont {Garcia}, \citenamefont {Kedves}, \citenamefont {Watanabe}, \citenamefont {Taniguchi}, \citenamefont {Sch\"onenberger},\ and\ \citenamefont {Makk}}]{Zihlmann2018}%
  \BibitemOpen
  \bibfield  {author} {\bibinfo {author} {\bibfnamefont {S.}~\bibnamefont {Zihlmann}}, \bibinfo {author} {\bibfnamefont {A.~W.}\ \bibnamefont {Cummings}}, \bibinfo {author} {\bibfnamefont {J.~H.}\ \bibnamefont {Garcia}}, \bibinfo {author} {\bibfnamefont {M.}~\bibnamefont {Kedves}}, \bibinfo {author} {\bibfnamefont {K.}~\bibnamefont {Watanabe}}, \bibinfo {author} {\bibfnamefont {T.}~\bibnamefont {Taniguchi}}, \bibinfo {author} {\bibfnamefont {C.}~\bibnamefont {Sch\"onenberger}},\ and\ \bibinfo {author} {\bibfnamefont {P.}~\bibnamefont {Makk}},\ }\bibfield  {title} {\bibinfo {title} {Large spin relaxation anisotropy and valley-zeeman spin-orbit coupling in ${\mathrm{wse}}_{2}$/graphene/$h$-bn heterostructures},\ }\href {https://doi.org/10.1103/PhysRevB.97.075434} {\bibfield  {journal} {\bibinfo  {journal} {Phys. Rev. B}\ }\textbf {\bibinfo {volume} {97}},\ \bibinfo {pages} {075434} (\bibinfo {year} {2018})}\BibitemShut {NoStop}%
\bibitem [{\citenamefont {Wakamura}\ \emph {et~al.}(2018)\citenamefont {Wakamura}, \citenamefont {Reale}, \citenamefont {Palczynski}, \citenamefont {Gu\'eron}, \citenamefont {Mattevi},\ and\ \citenamefont {Bouchiat}}]{Wakamura2018}%
  \BibitemOpen
  \bibfield  {author} {\bibinfo {author} {\bibfnamefont {T.}~\bibnamefont {Wakamura}}, \bibinfo {author} {\bibfnamefont {F.}~\bibnamefont {Reale}}, \bibinfo {author} {\bibfnamefont {P.}~\bibnamefont {Palczynski}}, \bibinfo {author} {\bibfnamefont {S.}~\bibnamefont {Gu\'eron}}, \bibinfo {author} {\bibfnamefont {C.}~\bibnamefont {Mattevi}},\ and\ \bibinfo {author} {\bibfnamefont {H.}~\bibnamefont {Bouchiat}},\ }\bibfield  {title} {\bibinfo {title} {Strong anisotropic spin-orbit interaction induced in graphene by monolayer ${\mathrm{ws}}_{2}$},\ }\href {https://doi.org/10.1103/PhysRevLett.120.106802} {\bibfield  {journal} {\bibinfo  {journal} {Phys. Rev. Lett.}\ }\textbf {\bibinfo {volume} {120}},\ \bibinfo {pages} {106802} (\bibinfo {year} {2018})}\BibitemShut {NoStop}%
\bibitem [{\citenamefont {Ghiasi}\ \emph {et~al.}(2019)\citenamefont {Ghiasi}, \citenamefont {Kaverzin}, \citenamefont {Blah},\ and\ \citenamefont {van Wees}}]{Ghiasi2019}%
  \BibitemOpen
  \bibfield  {author} {\bibinfo {author} {\bibfnamefont {T.~S.}\ \bibnamefont {Ghiasi}}, \bibinfo {author} {\bibfnamefont {A.~A.}\ \bibnamefont {Kaverzin}}, \bibinfo {author} {\bibfnamefont {P.~J.}\ \bibnamefont {Blah}},\ and\ \bibinfo {author} {\bibfnamefont {B.~J.}\ \bibnamefont {van Wees}},\ }\bibfield  {title} {\bibinfo {title} {Charge-to-spin conversion by the rashba{\textendash}edelstein effect in two-dimensional van der waals heterostructures up to room temperature},\ }\href {https://doi.org/10.1021/acs.nanolett.9b01611} {\bibfield  {journal} {\bibinfo  {journal} {Nano Letters}\ }\textbf {\bibinfo {volume} {19}},\ \bibinfo {pages} {5959} (\bibinfo {year} {2019})}\BibitemShut {NoStop}%
\bibitem [{\citenamefont {Island}\ \emph {et~al.}(2019)\citenamefont {Island}, \citenamefont {Cui}, \citenamefont {Lewandowski}, \citenamefont {Khoo}, \citenamefont {Spanton}, \citenamefont {Zhou}, \citenamefont {Rhodes}, \citenamefont {Hone}, \citenamefont {Taniguchi}, \citenamefont {Watanabe}, \citenamefont {Levitov}, \citenamefont {Zaletel},\ and\ \citenamefont {Young}}]{Island2019}%
  \BibitemOpen
  \bibfield  {author} {\bibinfo {author} {\bibfnamefont {J.~O.}\ \bibnamefont {Island}}, \bibinfo {author} {\bibfnamefont {X.}~\bibnamefont {Cui}}, \bibinfo {author} {\bibfnamefont {C.}~\bibnamefont {Lewandowski}}, \bibinfo {author} {\bibfnamefont {J.~Y.}\ \bibnamefont {Khoo}}, \bibinfo {author} {\bibfnamefont {E.~M.}\ \bibnamefont {Spanton}}, \bibinfo {author} {\bibfnamefont {H.}~\bibnamefont {Zhou}}, \bibinfo {author} {\bibfnamefont {D.}~\bibnamefont {Rhodes}}, \bibinfo {author} {\bibfnamefont {J.~C.}\ \bibnamefont {Hone}}, \bibinfo {author} {\bibfnamefont {T.}~\bibnamefont {Taniguchi}}, \bibinfo {author} {\bibfnamefont {K.}~\bibnamefont {Watanabe}}, \bibinfo {author} {\bibfnamefont {L.~S.}\ \bibnamefont {Levitov}}, \bibinfo {author} {\bibfnamefont {M.~P.}\ \bibnamefont {Zaletel}},\ and\ \bibinfo {author} {\bibfnamefont {A.~F.}\ \bibnamefont {Young}},\ }\bibfield  {title} {\bibinfo {title} {Spinorbit-driven band inversion in bilayer graphene by the van der waals proximity effect},\ }\href
  {https://doi.org/10.1038/s41586-019-1304-2} {\bibfield  {journal} {\bibinfo  {journal} {Nature}\ }\textbf {\bibinfo {volume} {571}},\ \bibinfo {pages} {85} (\bibinfo {year} {2019})}\BibitemShut {NoStop}%
\bibitem [{\citenamefont {Wang}\ \emph {et~al.}(2019)\citenamefont {Wang}, \citenamefont {Che}, \citenamefont {Cao}, \citenamefont {Lyu}, \citenamefont {Watanabe}, \citenamefont {Taniguchi}, \citenamefont {Lau},\ and\ \citenamefont {Bockrath}}]{Wang2019}%
  \BibitemOpen
  \bibfield  {author} {\bibinfo {author} {\bibfnamefont {D.}~\bibnamefont {Wang}}, \bibinfo {author} {\bibfnamefont {S.}~\bibnamefont {Che}}, \bibinfo {author} {\bibfnamefont {G.}~\bibnamefont {Cao}}, \bibinfo {author} {\bibfnamefont {R.}~\bibnamefont {Lyu}}, \bibinfo {author} {\bibfnamefont {K.}~\bibnamefont {Watanabe}}, \bibinfo {author} {\bibfnamefont {T.}~\bibnamefont {Taniguchi}}, \bibinfo {author} {\bibfnamefont {C.~N.}\ \bibnamefont {Lau}},\ and\ \bibinfo {author} {\bibfnamefont {M.}~\bibnamefont {Bockrath}},\ }\bibfield  {title} {\bibinfo {title} {Quantum hall effect measurement of spin{\textendash}orbit coupling strengths in ultraclean bilayer graphene/wse$_2$ heterostructures},\ }\href {https://doi.org/10.1021/acs.nanolett.9b02445} {\bibfield  {journal} {\bibinfo  {journal} {Nano Letters}\ }\textbf {\bibinfo {volume} {19}},\ \bibinfo {pages} {7028} (\bibinfo {year} {2019})}\BibitemShut {NoStop}%
\bibitem [{\citenamefont {Omar}\ \emph {et~al.}(2019)\citenamefont {Omar}, \citenamefont {Madhushankar},\ and\ \citenamefont {van Wees}}]{Omar2019}%
  \BibitemOpen
  \bibfield  {author} {\bibinfo {author} {\bibfnamefont {S.}~\bibnamefont {Omar}}, \bibinfo {author} {\bibfnamefont {B.~N.}\ \bibnamefont {Madhushankar}},\ and\ \bibinfo {author} {\bibfnamefont {B.~J.}\ \bibnamefont {van Wees}},\ }\bibfield  {title} {\bibinfo {title} {Large spin-relaxation anisotropy in bilayer-graphene/${\mathrm{ws}}_{2}$ heterostructures},\ }\href {https://doi.org/10.1103/PhysRevB.100.155415} {\bibfield  {journal} {\bibinfo  {journal} {Phys. Rev. B}\ }\textbf {\bibinfo {volume} {100}},\ \bibinfo {pages} {155415} (\bibinfo {year} {2019})}\BibitemShut {NoStop}%
\bibitem [{\citenamefont {Tiwari}\ \emph {et~al.}(2021)\citenamefont {Tiwari}, \citenamefont {Srivastav},\ and\ \citenamefont {Bid}}]{Tiwari2021}%
  \BibitemOpen
  \bibfield  {author} {\bibinfo {author} {\bibfnamefont {P.}~\bibnamefont {Tiwari}}, \bibinfo {author} {\bibfnamefont {S.~K.}\ \bibnamefont {Srivastav}},\ and\ \bibinfo {author} {\bibfnamefont {A.}~\bibnamefont {Bid}},\ }\bibfield  {title} {\bibinfo {title} {Electric-field-tunable valley zeeman effect in bilayer graphene heterostructures: Realization of the spin-orbit valve effect},\ }\href {https://doi.org/10.1103/PhysRevLett.126.096801} {\bibfield  {journal} {\bibinfo  {journal} {Phys. Rev. Lett.}\ }\textbf {\bibinfo {volume} {126}},\ \bibinfo {pages} {096801} (\bibinfo {year} {2021})}\BibitemShut {NoStop}%
\bibitem [{\citenamefont {Ingla-Ayn\'es}\ \emph {et~al.}(2021)\citenamefont {Ingla-Ayn\'es}, \citenamefont {Herling}, \citenamefont {Fabian}, \citenamefont {Hueso},\ and\ \citenamefont {Casanova}}]{InglaAynes2021}%
  \BibitemOpen
  \bibfield  {author} {\bibinfo {author} {\bibfnamefont {J.}~\bibnamefont {Ingla-Ayn\'es}}, \bibinfo {author} {\bibfnamefont {F.}~\bibnamefont {Herling}}, \bibinfo {author} {\bibfnamefont {J.}~\bibnamefont {Fabian}}, \bibinfo {author} {\bibfnamefont {L.~E.}\ \bibnamefont {Hueso}},\ and\ \bibinfo {author} {\bibfnamefont {F.}~\bibnamefont {Casanova}},\ }\bibfield  {title} {\bibinfo {title} {Electrical control of valley-zeeman spin-orbit-coupling--induced spin precession at room temperature},\ }\href {https://doi.org/10.1103/PhysRevLett.127.047202} {\bibfield  {journal} {\bibinfo  {journal} {Phys. Rev. Lett.}\ }\textbf {\bibinfo {volume} {127}},\ \bibinfo {pages} {047202} (\bibinfo {year} {2021})}\BibitemShut {NoStop}%
\bibitem [{\citenamefont {Amann}\ \emph {et~al.}(2022)\citenamefont {Amann}, \citenamefont {V\"olkl}, \citenamefont {Rockinger}, \citenamefont {Kochan}, \citenamefont {Watanabe}, \citenamefont {Taniguchi}, \citenamefont {Fabian}, \citenamefont {Weiss},\ and\ \citenamefont {Eroms}}]{Amann2022}%
  \BibitemOpen
  \bibfield  {author} {\bibinfo {author} {\bibfnamefont {J.}~\bibnamefont {Amann}}, \bibinfo {author} {\bibfnamefont {T.}~\bibnamefont {V\"olkl}}, \bibinfo {author} {\bibfnamefont {T.}~\bibnamefont {Rockinger}}, \bibinfo {author} {\bibfnamefont {D.}~\bibnamefont {Kochan}}, \bibinfo {author} {\bibfnamefont {K.}~\bibnamefont {Watanabe}}, \bibinfo {author} {\bibfnamefont {T.}~\bibnamefont {Taniguchi}}, \bibinfo {author} {\bibfnamefont {J.}~\bibnamefont {Fabian}}, \bibinfo {author} {\bibfnamefont {D.}~\bibnamefont {Weiss}},\ and\ \bibinfo {author} {\bibfnamefont {J.}~\bibnamefont {Eroms}},\ }\bibfield  {title} {\bibinfo {title} {Counterintuitive gate dependence of weak antilocalization in bilayer $\mathrm{graphene}/{\mathrm{wse}}_{2}$ heterostructures},\ }\href {https://doi.org/10.1103/PhysRevB.105.115425} {\bibfield  {journal} {\bibinfo  {journal} {Phys. Rev. B}\ }\textbf {\bibinfo {volume} {105}},\ \bibinfo {pages} {115425} (\bibinfo {year} {2022})}\BibitemShut {NoStop}%
\bibitem [{\citenamefont {Ingla-Ayn\'es}\ \emph {et~al.}(2015)\citenamefont {Ingla-Ayn\'es}, \citenamefont {Guimar\~aes}, \citenamefont {Meijerink}, \citenamefont {Zomer},\ and\ \citenamefont {van Wees}}]{InglaAynes2015}%
  \BibitemOpen
  \bibfield  {author} {\bibinfo {author} {\bibfnamefont {J.}~\bibnamefont {Ingla-Ayn\'es}}, \bibinfo {author} {\bibfnamefont {M.~H.~D.}\ \bibnamefont {Guimar\~aes}}, \bibinfo {author} {\bibfnamefont {R.~J.}\ \bibnamefont {Meijerink}}, \bibinfo {author} {\bibfnamefont {P.~J.}\ \bibnamefont {Zomer}},\ and\ \bibinfo {author} {\bibfnamefont {B.~J.}\ \bibnamefont {van Wees}},\ }\bibfield  {title} {\bibinfo {title} {$24\ensuremath{-}\ensuremath{\mu}\mathrm{m}$ spin relaxation length in boron nitride encapsulated bilayer graphene},\ }\href {https://doi.org/10.1103/PhysRevB.92.201410} {\bibfield  {journal} {\bibinfo  {journal} {Phys. Rev. B}\ }\textbf {\bibinfo {volume} {92}},\ \bibinfo {pages} {201410} (\bibinfo {year} {2015})}\BibitemShut {NoStop}%
\bibitem [{\citenamefont {Drögeler}\ \emph {et~al.}(2016)\citenamefont {Drögeler}, \citenamefont {Franzen}, \citenamefont {Volmer}, \citenamefont {Pohlmann}, \citenamefont {Banszerus}, \citenamefont {Wolter}, \citenamefont {Watanabe}, \citenamefont {Taniguchi}, \citenamefont {Stampfer},\ and\ \citenamefont {Beschoten}}]{Droegeler2016}%
  \BibitemOpen
  \bibfield  {author} {\bibinfo {author} {\bibfnamefont {M.}~\bibnamefont {Drögeler}}, \bibinfo {author} {\bibfnamefont {C.}~\bibnamefont {Franzen}}, \bibinfo {author} {\bibfnamefont {F.}~\bibnamefont {Volmer}}, \bibinfo {author} {\bibfnamefont {T.}~\bibnamefont {Pohlmann}}, \bibinfo {author} {\bibfnamefont {L.}~\bibnamefont {Banszerus}}, \bibinfo {author} {\bibfnamefont {M.}~\bibnamefont {Wolter}}, \bibinfo {author} {\bibfnamefont {K.}~\bibnamefont {Watanabe}}, \bibinfo {author} {\bibfnamefont {T.}~\bibnamefont {Taniguchi}}, \bibinfo {author} {\bibfnamefont {C.}~\bibnamefont {Stampfer}},\ and\ \bibinfo {author} {\bibfnamefont {B.}~\bibnamefont {Beschoten}},\ }\bibfield  {title} {\bibinfo {title} {Spin lifetimes exceeding 12 ns in graphene nonlocal spin valve devices},\ }\href {https://doi.org/10.1021/acs.nanolett.6b00497} {\bibfield  {journal} {\bibinfo  {journal} {Nano Letters}\ }\textbf {\bibinfo {volume} {16}},\ \bibinfo {pages} {3533} (\bibinfo {year} {2016})}\BibitemShut {NoStop}%
\bibitem [{\citenamefont {Singh}\ \emph {et~al.}(2016)\citenamefont {Singh}, \citenamefont {Katoch}, \citenamefont {Xu}, \citenamefont {Tan}, \citenamefont {Zhu}, \citenamefont {Amamou}, \citenamefont {Hone},\ and\ \citenamefont {Kawakami}}]{Singh2016}%
  \BibitemOpen
  \bibfield  {author} {\bibinfo {author} {\bibfnamefont {S.}~\bibnamefont {Singh}}, \bibinfo {author} {\bibfnamefont {J.}~\bibnamefont {Katoch}}, \bibinfo {author} {\bibfnamefont {J.}~\bibnamefont {Xu}}, \bibinfo {author} {\bibfnamefont {C.}~\bibnamefont {Tan}}, \bibinfo {author} {\bibfnamefont {T.}~\bibnamefont {Zhu}}, \bibinfo {author} {\bibfnamefont {W.}~\bibnamefont {Amamou}}, \bibinfo {author} {\bibfnamefont {J.}~\bibnamefont {Hone}},\ and\ \bibinfo {author} {\bibfnamefont {R.}~\bibnamefont {Kawakami}},\ }\bibfield  {title} {\bibinfo {title} {Nanosecond spin relaxation times in single layer graphene spin valves with hexagonal boron nitride tunnel barriers},\ }\href {https://doi.org/10.1063/1.4962635} {\bibfield  {journal} {\bibinfo  {journal} {Applied Physics Letters}\ }\textbf {\bibinfo {volume} {109}},\ \bibinfo {pages} {122411} (\bibinfo {year} {2016})}\BibitemShut {NoStop}%
\bibitem [{\citenamefont {Yang}\ \emph {et~al.}(2016)\citenamefont {Yang}, \citenamefont {Tu}, \citenamefont {Kim}, \citenamefont {Wu}, \citenamefont {Wang}, \citenamefont {Alicea}, \citenamefont {Wu}, \citenamefont {Bockrath},\ and\ \citenamefont {Shi}}]{Yang2016}%
  \BibitemOpen
  \bibfield  {author} {\bibinfo {author} {\bibfnamefont {B.}~\bibnamefont {Yang}}, \bibinfo {author} {\bibfnamefont {M.-F.}\ \bibnamefont {Tu}}, \bibinfo {author} {\bibfnamefont {J.}~\bibnamefont {Kim}}, \bibinfo {author} {\bibfnamefont {Y.}~\bibnamefont {Wu}}, \bibinfo {author} {\bibfnamefont {H.}~\bibnamefont {Wang}}, \bibinfo {author} {\bibfnamefont {J.}~\bibnamefont {Alicea}}, \bibinfo {author} {\bibfnamefont {R.}~\bibnamefont {Wu}}, \bibinfo {author} {\bibfnamefont {M.}~\bibnamefont {Bockrath}},\ and\ \bibinfo {author} {\bibfnamefont {J.}~\bibnamefont {Shi}},\ }\bibfield  {title} {\bibinfo {title} {Tunable spin–orbit coupling and symmetry-protected edge states in graphene/ws$_2$},\ }\href {https://doi.org/10.1088/2053-1583/3/3/031012} {\bibfield  {journal} {\bibinfo  {journal} {2D Materials}\ }\textbf {\bibinfo {volume} {3}},\ \bibinfo {pages} {031012} (\bibinfo {year} {2016})}\BibitemShut {NoStop}%
\bibitem [{\citenamefont {Dankert}\ and\ \citenamefont {Dash}(2017)}]{Dankert2017}%
  \BibitemOpen
  \bibfield  {author} {\bibinfo {author} {\bibfnamefont {A.}~\bibnamefont {Dankert}}\ and\ \bibinfo {author} {\bibfnamefont {S.~P.}\ \bibnamefont {Dash}},\ }\bibfield  {title} {\bibinfo {title} {Electrical gate control of spin current in van der waals heterostructures at room temperature},\ }\href {https://doi.org/10.1038/ncomms16093} {\bibfield  {journal} {\bibinfo  {journal} {Nature Communications}\ }\textbf {\bibinfo {volume} {8}},\ \bibinfo {pages} {16093} (\bibinfo {year} {2017})}\BibitemShut {NoStop}%
\bibitem [{\citenamefont {Omar}\ and\ \citenamefont {van Wees}(2018)}]{Omar2018}%
  \BibitemOpen
  \bibfield  {author} {\bibinfo {author} {\bibfnamefont {S.}~\bibnamefont {Omar}}\ and\ \bibinfo {author} {\bibfnamefont {B.~J.}\ \bibnamefont {van Wees}},\ }\bibfield  {title} {\bibinfo {title} {Spin transport in high-mobility graphene on ${\mathrm{ws}}_{2}$ substrate with electric-field tunable proximity spin-orbit interaction},\ }\href {https://doi.org/10.1103/PhysRevB.97.045414} {\bibfield  {journal} {\bibinfo  {journal} {Phys. Rev. B}\ }\textbf {\bibinfo {volume} {97}},\ \bibinfo {pages} {045414} (\bibinfo {year} {2018})}\BibitemShut {NoStop}%
\bibitem [{\citenamefont {Garcia}\ \emph {et~al.}(2017)\citenamefont {Garcia}, \citenamefont {Cummings},\ and\ \citenamefont {Roche}}]{Garcia2017}%
  \BibitemOpen
  \bibfield  {author} {\bibinfo {author} {\bibfnamefont {J.~H.}\ \bibnamefont {Garcia}}, \bibinfo {author} {\bibfnamefont {A.~W.}\ \bibnamefont {Cummings}},\ and\ \bibinfo {author} {\bibfnamefont {S.}~\bibnamefont {Roche}},\ }\bibfield  {title} {\bibinfo {title} {Spin hall effect and weak antilocalization in graphene/transition metal dichalcogenide heterostructures},\ }\href {https://doi.org/10.1021/acs.nanolett.7b02364} {\bibfield  {journal} {\bibinfo  {journal} {Nano Letters}\ }\textbf {\bibinfo {volume} {17}},\ \bibinfo {pages} {5078} (\bibinfo {year} {2017})}\BibitemShut {NoStop}%
\bibitem [{\citenamefont {Kane}\ and\ \citenamefont {Mele}(2005)}]{Kane2005}%
  \BibitemOpen
  \bibfield  {author} {\bibinfo {author} {\bibfnamefont {C.~L.}\ \bibnamefont {Kane}}\ and\ \bibinfo {author} {\bibfnamefont {E.~J.}\ \bibnamefont {Mele}},\ }\bibfield  {title} {\bibinfo {title} {Quantum spin hall effect in graphene},\ }\href {https://doi.org/10.1103/physrevlett.95.226801} {\bibfield  {journal} {\bibinfo  {journal} {Physical Review Letters}\ }\textbf {\bibinfo {volume} {95}},\ \bibinfo {pages} {226801} (\bibinfo {year} {2005})}\BibitemShut {NoStop}%
\bibitem [{\citenamefont {Gmitra}\ and\ \citenamefont {Fabian}(2017)}]{Gmitra2017}%
  \BibitemOpen
  \bibfield  {author} {\bibinfo {author} {\bibfnamefont {M.}~\bibnamefont {Gmitra}}\ and\ \bibinfo {author} {\bibfnamefont {J.}~\bibnamefont {Fabian}},\ }\bibfield  {title} {\bibinfo {title} {Proximity effects in bilayer graphene on monolayer ${\mathrm{wse}}_{2}$: Field-effect spin valley locking, spin-orbit valve, and spin transistor},\ }\href {https://doi.org/10.1103/PhysRevLett.119.146401} {\bibfield  {journal} {\bibinfo  {journal} {Phys. Rev. Lett.}\ }\textbf {\bibinfo {volume} {119}},\ \bibinfo {pages} {146401} (\bibinfo {year} {2017})}\BibitemShut {NoStop}%
\bibitem [{\citenamefont {David}\ \emph {et~al.}(2019)\citenamefont {David}, \citenamefont {Rakyta}, \citenamefont {Korm\'anyos},\ and\ \citenamefont {Burkard}}]{David2019}%
  \BibitemOpen
  \bibfield  {author} {\bibinfo {author} {\bibfnamefont {A.}~\bibnamefont {David}}, \bibinfo {author} {\bibfnamefont {P.}~\bibnamefont {Rakyta}}, \bibinfo {author} {\bibfnamefont {A.}~\bibnamefont {Korm\'anyos}},\ and\ \bibinfo {author} {\bibfnamefont {G.}~\bibnamefont {Burkard}},\ }\bibfield  {title} {\bibinfo {title} {Induced spin-orbit coupling in twisted graphene--transition metal dichalcogenide heterobilayers: Twistronics meets spintronics},\ }\href {https://doi.org/10.1103/PhysRevB.100.085412} {\bibfield  {journal} {\bibinfo  {journal} {Phys. Rev. B}\ }\textbf {\bibinfo {volume} {100}},\ \bibinfo {pages} {085412} (\bibinfo {year} {2019})}\BibitemShut {NoStop}%
\bibitem [{\citenamefont {Fülöp}\ \emph {et~al.}(2021{\natexlab{a}})\citenamefont {Fülöp}, \citenamefont {Márffy}, \citenamefont {Zihlmann}, \citenamefont {Gmitra}, \citenamefont {Tóvári}, \citenamefont {Szentpéteri}, \citenamefont {Kedves}, \citenamefont {Watanabe}, \citenamefont {Taniguchi}, \citenamefont {Fabian}, \citenamefont {Schönenberger}, \citenamefont {Makk},\ and\ \citenamefont {Csonka}}]{Fueloep2021}%
  \BibitemOpen
  \bibfield  {author} {\bibinfo {author} {\bibfnamefont {B.}~\bibnamefont {Fülöp}}, \bibinfo {author} {\bibfnamefont {A.}~\bibnamefont {Márffy}}, \bibinfo {author} {\bibfnamefont {S.}~\bibnamefont {Zihlmann}}, \bibinfo {author} {\bibfnamefont {M.}~\bibnamefont {Gmitra}}, \bibinfo {author} {\bibfnamefont {E.}~\bibnamefont {Tóvári}}, \bibinfo {author} {\bibfnamefont {B.}~\bibnamefont {Szentpéteri}}, \bibinfo {author} {\bibfnamefont {M.}~\bibnamefont {Kedves}}, \bibinfo {author} {\bibfnamefont {K.}~\bibnamefont {Watanabe}}, \bibinfo {author} {\bibfnamefont {T.}~\bibnamefont {Taniguchi}}, \bibinfo {author} {\bibfnamefont {J.}~\bibnamefont {Fabian}}, \bibinfo {author} {\bibfnamefont {C.}~\bibnamefont {Schönenberger}}, \bibinfo {author} {\bibfnamefont {P.}~\bibnamefont {Makk}},\ and\ \bibinfo {author} {\bibfnamefont {S.}~\bibnamefont {Csonka}},\ }\bibfield  {title} {\bibinfo {title} {Boosting proximity spin-orbit coupling in graphene/wse2 heterostructures via hydrostatic pressure},\ }\href
  {https://doi.org/10.1038/s41699-021-00262-9} {\bibfield  {journal} {\bibinfo  {journal} {npj 2D Materials and Applications}\ }\textbf {\bibinfo {volume} {5}},\ \bibinfo {pages} {82} (\bibinfo {year} {2021}{\natexlab{a}})}\BibitemShut {NoStop}%
\bibitem [{\citenamefont {Szentpéteri}\ \emph {et~al.}(2021)\citenamefont {Szentpéteri}, \citenamefont {Rickhaus}, \citenamefont {de~Vries}, \citenamefont {Márffy}, \citenamefont {Fülöp}, \citenamefont {Tóvári}, \citenamefont {Watanabe}, \citenamefont {Taniguchi}, \citenamefont {Kormányos}, \citenamefont {Csonka},\ and\ \citenamefont {Makk}}]{Szentpeteri2021}%
  \BibitemOpen
  \bibfield  {author} {\bibinfo {author} {\bibfnamefont {B.}~\bibnamefont {Szentpéteri}}, \bibinfo {author} {\bibfnamefont {P.}~\bibnamefont {Rickhaus}}, \bibinfo {author} {\bibfnamefont {F.~K.}\ \bibnamefont {de~Vries}}, \bibinfo {author} {\bibfnamefont {A.}~\bibnamefont {Márffy}}, \bibinfo {author} {\bibfnamefont {B.}~\bibnamefont {Fülöp}}, \bibinfo {author} {\bibfnamefont {E.}~\bibnamefont {Tóvári}}, \bibinfo {author} {\bibfnamefont {K.}~\bibnamefont {Watanabe}}, \bibinfo {author} {\bibfnamefont {T.}~\bibnamefont {Taniguchi}}, \bibinfo {author} {\bibfnamefont {A.}~\bibnamefont {Kormányos}}, \bibinfo {author} {\bibfnamefont {S.}~\bibnamefont {Csonka}},\ and\ \bibinfo {author} {\bibfnamefont {P.}~\bibnamefont {Makk}},\ }\bibfield  {title} {\bibinfo {title} {Tailoring the band structure of twisted double bilayer graphene with pressure},\ }\href {https://doi.org/10.1021/acs.nanolett.1c03066} {\bibfield  {journal} {\bibinfo  {journal} {Nano Letters}\ }\textbf {\bibinfo {volume} {21}},\ \bibinfo {pages}
  {8777} (\bibinfo {year} {2021})},\ \bibinfo {note} {pMID: 34662136},\ \Eprint {https://arxiv.org/abs/https://doi.org/10.1021/acs.nanolett.1c03066} {https://doi.org/10.1021/acs.nanolett.1c03066} \BibitemShut {NoStop}%
\bibitem [{\citenamefont {Fülöp}\ \emph {et~al.}(2021{\natexlab{b}})\citenamefont {Fülöp}, \citenamefont {Márffy}, \citenamefont {Tóvári}, \citenamefont {Kedves}, \citenamefont {Zihlmann}, \citenamefont {Indolese}, \citenamefont {Kovács-Krausz}, \citenamefont {Watanabe}, \citenamefont {Taniguchi}, \citenamefont {Schönenberger}, \citenamefont {Kézsmárki}, \citenamefont {Makk},\ and\ \citenamefont {Csonka}}]{Fueloep2021a}%
  \BibitemOpen
  \bibfield  {author} {\bibinfo {author} {\bibfnamefont {B.}~\bibnamefont {Fülöp}}, \bibinfo {author} {\bibfnamefont {A.}~\bibnamefont {Márffy}}, \bibinfo {author} {\bibfnamefont {E.}~\bibnamefont {Tóvári}}, \bibinfo {author} {\bibfnamefont {M.}~\bibnamefont {Kedves}}, \bibinfo {author} {\bibfnamefont {S.}~\bibnamefont {Zihlmann}}, \bibinfo {author} {\bibfnamefont {D.}~\bibnamefont {Indolese}}, \bibinfo {author} {\bibfnamefont {Z.}~\bibnamefont {Kovács-Krausz}}, \bibinfo {author} {\bibfnamefont {K.}~\bibnamefont {Watanabe}}, \bibinfo {author} {\bibfnamefont {T.}~\bibnamefont {Taniguchi}}, \bibinfo {author} {\bibfnamefont {C.}~\bibnamefont {Schönenberger}}, \bibinfo {author} {\bibfnamefont {I.}~\bibnamefont {Kézsmárki}}, \bibinfo {author} {\bibfnamefont {P.}~\bibnamefont {Makk}},\ and\ \bibinfo {author} {\bibfnamefont {S.}~\bibnamefont {Csonka}},\ }\bibfield  {title} {\bibinfo {title} {New method of transport measurements on van der waals heterostructures under pressure},\ }\href
  {https://doi.org/10.1063/5.0058583} {\bibfield  {journal} {\bibinfo  {journal} {Journal of Applied Physics}\ }\textbf {\bibinfo {volume} {130}},\ \bibinfo {pages} {64303} (\bibinfo {year} {2021}{\natexlab{b}})}\BibitemShut {NoStop}%
\bibitem [{\citenamefont {McCann}(2006)}]{McCann2006a}%
  \BibitemOpen
  \bibfield  {author} {\bibinfo {author} {\bibfnamefont {E.}~\bibnamefont {McCann}},\ }\bibfield  {title} {\bibinfo {title} {Asymmetry gap in the electronic band structure of bilayer graphene},\ }\href {https://doi.org/10.1103/PhysRevB.74.161403} {\bibfield  {journal} {\bibinfo  {journal} {Phys. Rev. B}\ }\textbf {\bibinfo {volume} {74}},\ \bibinfo {pages} {161403} (\bibinfo {year} {2006})}\BibitemShut {NoStop}%
\bibitem [{\citenamefont {Castro}\ \emph {et~al.}(2007)\citenamefont {Castro}, \citenamefont {Novoselov}, \citenamefont {Morozov}, \citenamefont {Peres}, \citenamefont {dos Santos}, \citenamefont {Nilsson}, \citenamefont {Guinea}, \citenamefont {Geim},\ and\ \citenamefont {Neto}}]{Castro2007}%
  \BibitemOpen
  \bibfield  {author} {\bibinfo {author} {\bibfnamefont {E.~V.}\ \bibnamefont {Castro}}, \bibinfo {author} {\bibfnamefont {K.~S.}\ \bibnamefont {Novoselov}}, \bibinfo {author} {\bibfnamefont {S.~V.}\ \bibnamefont {Morozov}}, \bibinfo {author} {\bibfnamefont {N.~M.~R.}\ \bibnamefont {Peres}}, \bibinfo {author} {\bibfnamefont {J.~M. B.~L.}\ \bibnamefont {dos Santos}}, \bibinfo {author} {\bibfnamefont {J.}~\bibnamefont {Nilsson}}, \bibinfo {author} {\bibfnamefont {F.}~\bibnamefont {Guinea}}, \bibinfo {author} {\bibfnamefont {A.~K.}\ \bibnamefont {Geim}},\ and\ \bibinfo {author} {\bibfnamefont {A.~H.~C.}\ \bibnamefont {Neto}},\ }\bibfield  {title} {\bibinfo {title} {Biased bilayer graphene: Semiconductor with a gap tunable by the electric field effect},\ }\href {https://doi.org/10.1103/PhysRevLett.99.216802} {\bibfield  {journal} {\bibinfo  {journal} {Phys. Rev. Lett.}\ }\textbf {\bibinfo {volume} {99}},\ \bibinfo {pages} {216802} (\bibinfo {year} {2007})}\BibitemShut {NoStop}%
\bibitem [{\citenamefont {Peñaranda}\ \emph {et~al.}(2023)\citenamefont {Peñaranda}, \citenamefont {Aguado}, \citenamefont {Prada},\ and\ \citenamefont {San-Jose}}]{Penaranda2023}%
  \BibitemOpen
  \bibfield  {author} {\bibinfo {author} {\bibfnamefont {F.}~\bibnamefont {Peñaranda}}, \bibinfo {author} {\bibfnamefont {R.}~\bibnamefont {Aguado}}, \bibinfo {author} {\bibfnamefont {E.}~\bibnamefont {Prada}},\ and\ \bibinfo {author} {\bibfnamefont {P.}~\bibnamefont {San-Jose}},\ }\bibfield  {title} {\bibinfo {title} {{Majorana bound states in encapsulated bilayer graphene}},\ }\href {https://doi.org/10.21468/SciPostPhys.14.4.075} {\bibfield  {journal} {\bibinfo  {journal} {SciPost Phys.}\ }\textbf {\bibinfo {volume} {14}},\ \bibinfo {pages} {075} (\bibinfo {year} {2023})}\BibitemShut {NoStop}%
\bibitem [{\citenamefont {Zaletel}\ and\ \citenamefont {Khoo}()}]{Zaletel2019}%
  \BibitemOpen
  \bibfield  {author} {\bibinfo {author} {\bibfnamefont {M.~P.}\ \bibnamefont {Zaletel}}\ and\ \bibinfo {author} {\bibfnamefont {J.~Y.}\ \bibnamefont {Khoo}},\ }\bibfield  {title} {\bibinfo {title} {The gate-tunable strong and fragile topology of multilayer-graphene on a transition metal dichalcogenide},\ }\bibfield  {journal} {\bibinfo  {journal} {arXiv (Condensed Matter, Mesoscale and Nanoscale Physics)}\ }\href {https://doi.org/10.48550/ARXIV.1901.01294} {10.48550/ARXIV.1901.01294},\ \bibinfo {note} {january 9, 2019, \url{https://arxiv.org/abs/1901.01294} (accessed 2023-09-01)},\ \Eprint {https://arxiv.org/abs/1901.01294} {arXiv:1901.01294 [cond-mat.mes-hall]} \BibitemShut {NoStop}%
\bibitem [{\citenamefont {Yankowitz}\ \emph {et~al.}(2018)\citenamefont {Yankowitz}, \citenamefont {Jung}, \citenamefont {Laksono}, \citenamefont {Leconte}, \citenamefont {Chittari}, \citenamefont {Watanabe}, \citenamefont {Taniguchi}, \citenamefont {Adam}, \citenamefont {Graf},\ and\ \citenamefont {Dean}}]{Yankowitz2018}%
  \BibitemOpen
  \bibfield  {author} {\bibinfo {author} {\bibfnamefont {M.}~\bibnamefont {Yankowitz}}, \bibinfo {author} {\bibfnamefont {J.}~\bibnamefont {Jung}}, \bibinfo {author} {\bibfnamefont {E.}~\bibnamefont {Laksono}}, \bibinfo {author} {\bibfnamefont {N.}~\bibnamefont {Leconte}}, \bibinfo {author} {\bibfnamefont {B.~L.}\ \bibnamefont {Chittari}}, \bibinfo {author} {\bibfnamefont {K.}~\bibnamefont {Watanabe}}, \bibinfo {author} {\bibfnamefont {T.}~\bibnamefont {Taniguchi}}, \bibinfo {author} {\bibfnamefont {S.}~\bibnamefont {Adam}}, \bibinfo {author} {\bibfnamefont {D.}~\bibnamefont {Graf}},\ and\ \bibinfo {author} {\bibfnamefont {C.~R.}\ \bibnamefont {Dean}},\ }\bibfield  {title} {\bibinfo {title} {Dynamic band-structure tuning of graphene moiré superlattices with pressure},\ }\href {https://doi.org/10.1038/s41586-018-0107-1} {\bibfield  {journal} {\bibinfo  {journal} {Nature}\ }\textbf {\bibinfo {volume} {557}},\ \bibinfo {pages} {404} (\bibinfo {year} {2018})}\BibitemShut {NoStop}%
\bibitem [{\citenamefont {Carr}\ \emph {et~al.}(2018)\citenamefont {Carr}, \citenamefont {Fang}, \citenamefont {Jarillo-Herrero},\ and\ \citenamefont {Kaxiras}}]{Carr2018}%
  \BibitemOpen
  \bibfield  {author} {\bibinfo {author} {\bibfnamefont {S.}~\bibnamefont {Carr}}, \bibinfo {author} {\bibfnamefont {S.}~\bibnamefont {Fang}}, \bibinfo {author} {\bibfnamefont {P.}~\bibnamefont {Jarillo-Herrero}},\ and\ \bibinfo {author} {\bibfnamefont {E.}~\bibnamefont {Kaxiras}},\ }\bibfield  {title} {\bibinfo {title} {Pressure dependence of the magic twist angle in graphene superlattices},\ }\href {https://doi.org/10.1103/physrevb.98.085144} {\bibfield  {journal} {\bibinfo  {journal} {Physical Review B}\ }\textbf {\bibinfo {volume} {98}},\ \bibinfo {pages} {085144} (\bibinfo {year} {2018})}\BibitemShut {NoStop}%
\bibitem [{\citenamefont {Sui}\ \emph {et~al.}(2015)\citenamefont {Sui}, \citenamefont {Chen}, \citenamefont {Ma}, \citenamefont {Shan}, \citenamefont {Tian}, \citenamefont {Watanabe}, \citenamefont {Taniguchi}, \citenamefont {Jin}, \citenamefont {Yao}, \citenamefont {Xiao},\ and\ \citenamefont {Zhang}}]{Sui2015}%
  \BibitemOpen
  \bibfield  {author} {\bibinfo {author} {\bibfnamefont {M.}~\bibnamefont {Sui}}, \bibinfo {author} {\bibfnamefont {G.}~\bibnamefont {Chen}}, \bibinfo {author} {\bibfnamefont {L.}~\bibnamefont {Ma}}, \bibinfo {author} {\bibfnamefont {W.-Y.}\ \bibnamefont {Shan}}, \bibinfo {author} {\bibfnamefont {D.}~\bibnamefont {Tian}}, \bibinfo {author} {\bibfnamefont {K.}~\bibnamefont {Watanabe}}, \bibinfo {author} {\bibfnamefont {T.}~\bibnamefont {Taniguchi}}, \bibinfo {author} {\bibfnamefont {X.}~\bibnamefont {Jin}}, \bibinfo {author} {\bibfnamefont {W.}~\bibnamefont {Yao}}, \bibinfo {author} {\bibfnamefont {D.}~\bibnamefont {Xiao}},\ and\ \bibinfo {author} {\bibfnamefont {Y.}~\bibnamefont {Zhang}},\ }\bibfield  {title} {\bibinfo {title} {Gate-tunable topological valley transport in bilayer graphene},\ }\href {https://doi.org/10.1038/nphys3485} {\bibfield  {journal} {\bibinfo  {journal} {Nature Physics}\ }\textbf {\bibinfo {volume} {11}},\ \bibinfo {pages} {1027} (\bibinfo {year} {2015})}\BibitemShut {NoStop}%
\bibitem [{\citenamefont {McCann}\ and\ \citenamefont {Fal'ko}(2006)}]{McCann2006}%
  \BibitemOpen
  \bibfield  {author} {\bibinfo {author} {\bibfnamefont {E.}~\bibnamefont {McCann}}\ and\ \bibinfo {author} {\bibfnamefont {V.~I.}\ \bibnamefont {Fal'ko}},\ }\bibfield  {title} {\bibinfo {title} {Landau-level degeneracy and quantum hall effect in a graphite bilayer},\ }\href {https://doi.org/10.1103/PhysRevLett.96.086805} {\bibfield  {journal} {\bibinfo  {journal} {Phys. Rev. Lett.}\ }\textbf {\bibinfo {volume} {96}},\ \bibinfo {pages} {086805} (\bibinfo {year} {2006})}\BibitemShut {NoStop}%
\bibitem [{\citenamefont {Novoselov}\ \emph {et~al.}(2006)\citenamefont {Novoselov}, \citenamefont {McCann}, \citenamefont {Morozov}, \citenamefont {Fal'ko}, \citenamefont {Katsnelson}, \citenamefont {Zeitler}, \citenamefont {Jiang}, \citenamefont {Schedin},\ and\ \citenamefont {Geim}}]{Novoselov2006}%
  \BibitemOpen
  \bibfield  {author} {\bibinfo {author} {\bibfnamefont {K.~S.}\ \bibnamefont {Novoselov}}, \bibinfo {author} {\bibfnamefont {E.}~\bibnamefont {McCann}}, \bibinfo {author} {\bibfnamefont {S.~V.}\ \bibnamefont {Morozov}}, \bibinfo {author} {\bibfnamefont {V.~I.}\ \bibnamefont {Fal'ko}}, \bibinfo {author} {\bibfnamefont {M.~I.}\ \bibnamefont {Katsnelson}}, \bibinfo {author} {\bibfnamefont {U.}~\bibnamefont {Zeitler}}, \bibinfo {author} {\bibfnamefont {D.}~\bibnamefont {Jiang}}, \bibinfo {author} {\bibfnamefont {F.}~\bibnamefont {Schedin}},\ and\ \bibinfo {author} {\bibfnamefont {A.~K.}\ \bibnamefont {Geim}},\ }\bibfield  {title} {\bibinfo {title} {Unconventional quantum hall effect and berry's phase of 2$\pi$ in bilayer graphene},\ }\href {https://doi.org/10.1038/nphys245} {\bibfield  {journal} {\bibinfo  {journal} {Nature Physics}\ }\textbf {\bibinfo {volume} {2}},\ \bibinfo {pages} {177} (\bibinfo {year} {2006})}\BibitemShut {NoStop}%
\bibitem [{\citenamefont {Hunt}\ \emph {et~al.}(2017)\citenamefont {Hunt}, \citenamefont {Li}, \citenamefont {Zibrov}, \citenamefont {Wang}, \citenamefont {Taniguchi}, \citenamefont {Watanabe}, \citenamefont {Hone}, \citenamefont {Dean}, \citenamefont {Zaletel}, \citenamefont {Ashoori},\ and\ \citenamefont {Young}}]{Hunt2017}%
  \BibitemOpen
  \bibfield  {author} {\bibinfo {author} {\bibfnamefont {B.~M.}\ \bibnamefont {Hunt}}, \bibinfo {author} {\bibfnamefont {J.~I.~A.}\ \bibnamefont {Li}}, \bibinfo {author} {\bibfnamefont {A.~A.}\ \bibnamefont {Zibrov}}, \bibinfo {author} {\bibfnamefont {L.}~\bibnamefont {Wang}}, \bibinfo {author} {\bibfnamefont {T.}~\bibnamefont {Taniguchi}}, \bibinfo {author} {\bibfnamefont {K.}~\bibnamefont {Watanabe}}, \bibinfo {author} {\bibfnamefont {J.}~\bibnamefont {Hone}}, \bibinfo {author} {\bibfnamefont {C.~R.}\ \bibnamefont {Dean}}, \bibinfo {author} {\bibfnamefont {M.}~\bibnamefont {Zaletel}}, \bibinfo {author} {\bibfnamefont {R.~C.}\ \bibnamefont {Ashoori}},\ and\ \bibinfo {author} {\bibfnamefont {A.~F.}\ \bibnamefont {Young}},\ }\bibfield  {title} {\bibinfo {title} {Direct measurement of discrete valley and orbital quantum numbers in bilayer graphene},\ }\href {https://doi.org/10.1038/s41467-017-00824-w} {\bibfield  {journal} {\bibinfo  {journal} {Nature Communications}\ }\textbf {\bibinfo {volume} {8}},\ \bibinfo
  {pages} {948} (\bibinfo {year} {2017})}\BibitemShut {NoStop}%
\bibitem [{\citenamefont {Khoo}\ and\ \citenamefont {Levitov}(2018)}]{Khoo2018}%
  \BibitemOpen
  \bibfield  {author} {\bibinfo {author} {\bibfnamefont {J.~Y.}\ \bibnamefont {Khoo}}\ and\ \bibinfo {author} {\bibfnamefont {L.}~\bibnamefont {Levitov}},\ }\bibfield  {title} {\bibinfo {title} {Tunable quantum hall edge conduction in bilayer graphene through spin-orbit interaction},\ }\href {https://doi.org/10.1103/physrevb.98.115307} {\bibfield  {journal} {\bibinfo  {journal} {Physical Review B}\ }\textbf {\bibinfo {volume} {98}},\ \bibinfo {pages} {115307} (\bibinfo {year} {2018})}\BibitemShut {NoStop}%
\bibitem [{\citenamefont {Sanchez-Yamagishi}\ \emph {et~al.}(2016)\citenamefont {Sanchez-Yamagishi}, \citenamefont {Luo}, \citenamefont {Young}, \citenamefont {Hunt}, \citenamefont {Watanabe}, \citenamefont {Taniguchi}, \citenamefont {Ashoori},\ and\ \citenamefont {Jarillo-Herrero}}]{SanchezYamagishi2016}%
  \BibitemOpen
  \bibfield  {author} {\bibinfo {author} {\bibfnamefont {J.~D.}\ \bibnamefont {Sanchez-Yamagishi}}, \bibinfo {author} {\bibfnamefont {J.~Y.}\ \bibnamefont {Luo}}, \bibinfo {author} {\bibfnamefont {A.~F.}\ \bibnamefont {Young}}, \bibinfo {author} {\bibfnamefont {B.~M.}\ \bibnamefont {Hunt}}, \bibinfo {author} {\bibfnamefont {K.}~\bibnamefont {Watanabe}}, \bibinfo {author} {\bibfnamefont {T.}~\bibnamefont {Taniguchi}}, \bibinfo {author} {\bibfnamefont {R.~C.}\ \bibnamefont {Ashoori}},\ and\ \bibinfo {author} {\bibfnamefont {P.}~\bibnamefont {Jarillo-Herrero}},\ }\bibfield  {title} {\bibinfo {title} {Helical edge states and fractional quantum hall effect in a graphene electron{\textendash}hole bilayer},\ }\href {https://doi.org/10.1038/nnano.2016.214} {\bibfield  {journal} {\bibinfo  {journal} {Nature Nanotechnology}\ }\textbf {\bibinfo {volume} {12}},\ \bibinfo {pages} {118} (\bibinfo {year} {2016})}\BibitemShut {NoStop}%
\bibitem [{\citenamefont {Veyrat}\ \emph {et~al.}(2020)\citenamefont {Veyrat}, \citenamefont {D{\'{e}}prez}, \citenamefont {Coissard}, \citenamefont {Li}, \citenamefont {Gay}, \citenamefont {Watanabe}, \citenamefont {Taniguchi}, \citenamefont {Han}, \citenamefont {Piot}, \citenamefont {Sellier},\ and\ \citenamefont {Sac{\'{e}}p{\'{e}}}}]{Veyrat2020}%
  \BibitemOpen
  \bibfield  {author} {\bibinfo {author} {\bibfnamefont {L.}~\bibnamefont {Veyrat}}, \bibinfo {author} {\bibfnamefont {C.}~\bibnamefont {D{\'{e}}prez}}, \bibinfo {author} {\bibfnamefont {A.}~\bibnamefont {Coissard}}, \bibinfo {author} {\bibfnamefont {X.}~\bibnamefont {Li}}, \bibinfo {author} {\bibfnamefont {F.}~\bibnamefont {Gay}}, \bibinfo {author} {\bibfnamefont {K.}~\bibnamefont {Watanabe}}, \bibinfo {author} {\bibfnamefont {T.}~\bibnamefont {Taniguchi}}, \bibinfo {author} {\bibfnamefont {Z.}~\bibnamefont {Han}}, \bibinfo {author} {\bibfnamefont {B.~A.}\ \bibnamefont {Piot}}, \bibinfo {author} {\bibfnamefont {H.}~\bibnamefont {Sellier}},\ and\ \bibinfo {author} {\bibfnamefont {B.}~\bibnamefont {Sac{\'{e}}p{\'{e}}}},\ }\bibfield  {title} {\bibinfo {title} {Helical quantum hall phase in graphene on srtio$_3$},\ }\href {https://doi.org/10.1126/science.aax8201} {\bibfield  {journal} {\bibinfo  {journal} {Science}\ }\textbf {\bibinfo {volume} {367}},\ \bibinfo {pages} {781} (\bibinfo {year} {2020})}\BibitemShut
  {NoStop}%
\bibitem [{\citenamefont {Hart}\ \emph {et~al.}(2014)\citenamefont {Hart}, \citenamefont {Ren}, \citenamefont {Wagner}, \citenamefont {Leubner}, \citenamefont {Mühlbauer}, \citenamefont {Brüne}, \citenamefont {Buhmann}, \citenamefont {Molenkamp},\ and\ \citenamefont {Yacoby}}]{Hart2014}%
  \BibitemOpen
  \bibfield  {author} {\bibinfo {author} {\bibfnamefont {S.}~\bibnamefont {Hart}}, \bibinfo {author} {\bibfnamefont {H.}~\bibnamefont {Ren}}, \bibinfo {author} {\bibfnamefont {T.}~\bibnamefont {Wagner}}, \bibinfo {author} {\bibfnamefont {P.}~\bibnamefont {Leubner}}, \bibinfo {author} {\bibfnamefont {M.}~\bibnamefont {Mühlbauer}}, \bibinfo {author} {\bibfnamefont {C.}~\bibnamefont {Brüne}}, \bibinfo {author} {\bibfnamefont {H.}~\bibnamefont {Buhmann}}, \bibinfo {author} {\bibfnamefont {L.~W.}\ \bibnamefont {Molenkamp}},\ and\ \bibinfo {author} {\bibfnamefont {A.}~\bibnamefont {Yacoby}},\ }\bibfield  {title} {\bibinfo {title} {Induced superconductivity in the quantum spin hall edge},\ }\href {https://doi.org/10.1038/nphys3036} {\bibfield  {journal} {\bibinfo  {journal} {Nature Physics}\ }\textbf {\bibinfo {volume} {10}},\ \bibinfo {pages} {638} (\bibinfo {year} {2014})}\BibitemShut {NoStop}%
\bibitem [{\citenamefont {Indolese}\ \emph {et~al.}(2018)\citenamefont {Indolese}, \citenamefont {Delagrange}, \citenamefont {Makk}, \citenamefont {Wallbank}, \citenamefont {Wanatabe}, \citenamefont {Taniguchi},\ and\ \citenamefont {Sch\"onenberger}}]{Indolese2018}%
  \BibitemOpen
  \bibfield  {author} {\bibinfo {author} {\bibfnamefont {D.~I.}\ \bibnamefont {Indolese}}, \bibinfo {author} {\bibfnamefont {R.}~\bibnamefont {Delagrange}}, \bibinfo {author} {\bibfnamefont {P.}~\bibnamefont {Makk}}, \bibinfo {author} {\bibfnamefont {J.~R.}\ \bibnamefont {Wallbank}}, \bibinfo {author} {\bibfnamefont {K.}~\bibnamefont {Wanatabe}}, \bibinfo {author} {\bibfnamefont {T.}~\bibnamefont {Taniguchi}},\ and\ \bibinfo {author} {\bibfnamefont {C.}~\bibnamefont {Sch\"onenberger}},\ }\bibfield  {title} {\bibinfo {title} {Signatures of van hove singularities probed by the supercurrent in a graphene-hbn superlattice},\ }\href {https://doi.org/10.1103/PhysRevLett.121.137701} {\bibfield  {journal} {\bibinfo  {journal} {Phys. Rev. Lett.}\ }\textbf {\bibinfo {volume} {121}},\ \bibinfo {pages} {137701} (\bibinfo {year} {2018})}\BibitemShut {NoStop}%
\bibitem [{\citenamefont {Li}\ \emph {et~al.}(2012)\citenamefont {Li}, \citenamefont {Qiao}, \citenamefont {Jung},\ and\ \citenamefont {Niu}}]{Li2012}%
  \BibitemOpen
  \bibfield  {author} {\bibinfo {author} {\bibfnamefont {X.}~\bibnamefont {Li}}, \bibinfo {author} {\bibfnamefont {Z.}~\bibnamefont {Qiao}}, \bibinfo {author} {\bibfnamefont {J.}~\bibnamefont {Jung}},\ and\ \bibinfo {author} {\bibfnamefont {Q.}~\bibnamefont {Niu}},\ }\bibfield  {title} {\bibinfo {title} {Unbalanced edge modes and topological phase transition in gated trilayer graphene},\ }\href {https://doi.org/10.1103/PhysRevB.85.201404} {\bibfield  {journal} {\bibinfo  {journal} {Phys. Rev. B}\ }\textbf {\bibinfo {volume} {85}},\ \bibinfo {pages} {201404} (\bibinfo {year} {2012})}\BibitemShut {NoStop}%
\bibitem [{\citenamefont {Wang}\ \emph {et~al.}(2020)\citenamefont {Wang}, \citenamefont {Bultinck},\ and\ \citenamefont {Zaletel}}]{Wang2020}%
  \BibitemOpen
  \bibfield  {author} {\bibinfo {author} {\bibfnamefont {T.}~\bibnamefont {Wang}}, \bibinfo {author} {\bibfnamefont {N.}~\bibnamefont {Bultinck}},\ and\ \bibinfo {author} {\bibfnamefont {M.~P.}\ \bibnamefont {Zaletel}},\ }\bibfield  {title} {\bibinfo {title} {Flat-band topology of magic angle graphene on a transition metal dichalcogenide},\ }\href {https://doi.org/10.1103/physrevb.102.235146} {\bibfield  {journal} {\bibinfo  {journal} {Physical Review B}\ }\textbf {\bibinfo {volume} {102}},\ \bibinfo {pages} {235146} (\bibinfo {year} {2020})}\BibitemShut {NoStop}%
\end{thebibliography}%


\begin{thebibliography}{14}%
\makeatletter
\providecommand \@ifxundefined [1]{%
 \@ifx{#1\undefined}
}%
\providecommand \@ifnum [1]{%
 \ifnum #1\expandafter \@firstoftwo
 \else \expandafter \@secondoftwo
 \fi
}%
\providecommand \@ifx [1]{%
 \ifx #1\expandafter \@firstoftwo
 \else \expandafter \@secondoftwo
 \fi
}%
\providecommand \natexlab [1]{#1}%
\providecommand \enquote  [1]{``#1''}%
\providecommand \bibnamefont  [1]{#1}%
\providecommand \bibfnamefont [1]{#1}%
\providecommand \citenamefont [1]{#1}%
\providecommand \href@noop [0]{\@secondoftwo}%
\providecommand \href [0]{\begingroup \@sanitize@url \@href}%
\providecommand \@href[1]{\@@startlink{#1}\@@href}%
\providecommand \@@href[1]{\endgroup#1\@@endlink}%
\providecommand \@sanitize@url [0]{\catcode `\\12\catcode `\$12\catcode `\&12\catcode `\#12\catcode `\^12\catcode `\_12\catcode `\%12\relax}%
\providecommand \@@startlink[1]{}%
\providecommand \@@endlink[0]{}%
\providecommand \url  [0]{\begingroup\@sanitize@url \@url }%
\providecommand \@url [1]{\endgroup\@href {#1}{\urlprefix }}%
\providecommand \urlprefix  [0]{URL }%
\providecommand \Eprint [0]{\href }%
\providecommand \doibase [0]{https://doi.org/}%
\providecommand \selectlanguage [0]{\@gobble}%
\providecommand \bibinfo  [0]{\@secondoftwo}%
\providecommand \bibfield  [0]{\@secondoftwo}%
\providecommand \translation [1]{[#1]}%
\providecommand \BibitemOpen [0]{}%
\providecommand \bibitemStop [0]{}%
\providecommand \bibitemNoStop [0]{.\EOS\space}%
\providecommand \EOS [0]{\spacefactor3000\relax}%
\providecommand \BibitemShut  [1]{\csname bibitem#1\endcsname}%
\let\auto@bib@innerbib\@empty
%</preamble>
\bibitem [{\citenamefont {Fülöp}\ \emph {et~al.}(2021{\natexlab{a}})\citenamefont {Fülöp}, \citenamefont {Márffy}, \citenamefont {Tóvári}, \citenamefont {Kedves}, \citenamefont {Zihlmann}, \citenamefont {Indolese}, \citenamefont {Kovács-Krausz}, \citenamefont {Watanabe}, \citenamefont {Taniguchi}, \citenamefont {Schönenberger}, \citenamefont {Kézsmárki}, \citenamefont {Makk},\ and\ \citenamefont {Csonka}}]{Fueloep2021a}%
  \BibitemOpen
  \bibfield  {author} {\bibinfo {author} {\bibfnamefont {B.}~\bibnamefont {Fülöp}}, \bibinfo {author} {\bibfnamefont {A.}~\bibnamefont {Márffy}}, \bibinfo {author} {\bibfnamefont {E.}~\bibnamefont {Tóvári}}, \bibinfo {author} {\bibfnamefont {M.}~\bibnamefont {Kedves}}, \bibinfo {author} {\bibfnamefont {S.}~\bibnamefont {Zihlmann}}, \bibinfo {author} {\bibfnamefont {D.}~\bibnamefont {Indolese}}, \bibinfo {author} {\bibfnamefont {Z.}~\bibnamefont {Kovács-Krausz}}, \bibinfo {author} {\bibfnamefont {K.}~\bibnamefont {Watanabe}}, \bibinfo {author} {\bibfnamefont {T.}~\bibnamefont {Taniguchi}}, \bibinfo {author} {\bibfnamefont {C.}~\bibnamefont {Schönenberger}}, \bibinfo {author} {\bibfnamefont {I.}~\bibnamefont {Kézsmárki}}, \bibinfo {author} {\bibfnamefont {P.}~\bibnamefont {Makk}},\ and\ \bibinfo {author} {\bibfnamefont {S.}~\bibnamefont {Csonka}},\ }\bibfield  {title} {\bibinfo {title} {New method of transport measurements on van der waals heterostructures under pressure},\ }\href
  {https://doi.org/10.1063/5.0058583} {\bibfield  {journal} {\bibinfo  {journal} {Journal of Applied Physics}\ }\textbf {\bibinfo {volume} {130}},\ \bibinfo {pages} {64303} (\bibinfo {year} {2021}{\natexlab{a}})}\BibitemShut {NoStop}%
\bibitem [{\citenamefont {McCann}\ and\ \citenamefont {Koshino}(2013)}]{McCann2013}%
  \BibitemOpen
  \bibfield  {author} {\bibinfo {author} {\bibfnamefont {E.}~\bibnamefont {McCann}}\ and\ \bibinfo {author} {\bibfnamefont {M.}~\bibnamefont {Koshino}},\ }\bibfield  {title} {\bibinfo {title} {The electronic properties of bilayer graphene},\ }\href {https://doi.org/10.1088/0034-4885/76/5/056503} {\bibfield  {journal} {\bibinfo  {journal} {Reports on Progress in Physics}\ }\textbf {\bibinfo {volume} {76}},\ \bibinfo {pages} {056503} (\bibinfo {year} {2013})}\BibitemShut {NoStop}%
\bibitem [{\citenamefont {Konschuh}\ \emph {et~al.}(2012)\citenamefont {Konschuh}, \citenamefont {Gmitra}, \citenamefont {Kochan},\ and\ \citenamefont {Fabian}}]{Konschuh2012}%
  \BibitemOpen
  \bibfield  {author} {\bibinfo {author} {\bibfnamefont {S.}~\bibnamefont {Konschuh}}, \bibinfo {author} {\bibfnamefont {M.}~\bibnamefont {Gmitra}}, \bibinfo {author} {\bibfnamefont {D.}~\bibnamefont {Kochan}},\ and\ \bibinfo {author} {\bibfnamefont {J.}~\bibnamefont {Fabian}},\ }\bibfield  {title} {\bibinfo {title} {Theory of spin-orbit coupling in bilayer graphene},\ }\href {https://doi.org/10.1103/PhysRevB.85.115423} {\bibfield  {journal} {\bibinfo  {journal} {Phys. Rev. B}\ }\textbf {\bibinfo {volume} {85}},\ \bibinfo {pages} {115423} (\bibinfo {year} {2012})}\BibitemShut {NoStop}%
\bibitem [{\citenamefont {Zollner}\ and\ \citenamefont {Fabian}(2021)}]{Zollner2021}%
  \BibitemOpen
  \bibfield  {author} {\bibinfo {author} {\bibfnamefont {K.}~\bibnamefont {Zollner}}\ and\ \bibinfo {author} {\bibfnamefont {J.}~\bibnamefont {Fabian}},\ }\bibfield  {title} {\bibinfo {title} {Bilayer graphene encapsulated within monolayers of ${\mathrm{ws}}_{2}$ or ${\mathrm{cr}}_{2}{\mathrm{ge}}_{2}{\mathrm{te}}_{6}$: Tunable proximity spin-orbit or exchange coupling},\ }\href {https://doi.org/10.1103/PhysRevB.104.075126} {\bibfield  {journal} {\bibinfo  {journal} {Phys. Rev. B}\ }\textbf {\bibinfo {volume} {104}},\ \bibinfo {pages} {075126} (\bibinfo {year} {2021})}\BibitemShut {NoStop}%
\bibitem [{\citenamefont {Jung}\ and\ \citenamefont {MacDonald}(2014)}]{Jung2014}%
  \BibitemOpen
  \bibfield  {author} {\bibinfo {author} {\bibfnamefont {J.}~\bibnamefont {Jung}}\ and\ \bibinfo {author} {\bibfnamefont {A.~H.}\ \bibnamefont {MacDonald}},\ }\bibfield  {title} {\bibinfo {title} {Accurate tight-binding models for the $\ensuremath{\pi}$ bands of bilayer graphene},\ }\href {https://doi.org/10.1103/PhysRevB.89.035405} {\bibfield  {journal} {\bibinfo  {journal} {Phys. Rev. B}\ }\textbf {\bibinfo {volume} {89}},\ \bibinfo {pages} {035405} (\bibinfo {year} {2014})}\BibitemShut {NoStop}%
\bibitem [{\citenamefont {Khoo}\ and\ \citenamefont {Levitov}(2018)}]{Khoo2018}%
  \BibitemOpen
  \bibfield  {author} {\bibinfo {author} {\bibfnamefont {J.~Y.}\ \bibnamefont {Khoo}}\ and\ \bibinfo {author} {\bibfnamefont {L.}~\bibnamefont {Levitov}},\ }\bibfield  {title} {\bibinfo {title} {Tunable quantum hall edge conduction in bilayer graphene through spin-orbit interaction},\ }\href {https://doi.org/10.1103/physrevb.98.115307} {\bibfield  {journal} {\bibinfo  {journal} {Physical Review B}\ }\textbf {\bibinfo {volume} {98}},\ \bibinfo {pages} {115307} (\bibinfo {year} {2018})}\BibitemShut {NoStop}%
\bibitem [{\citenamefont {Hunt}\ \emph {et~al.}(2017)\citenamefont {Hunt}, \citenamefont {Li}, \citenamefont {Zibrov}, \citenamefont {Wang}, \citenamefont {Taniguchi}, \citenamefont {Watanabe}, \citenamefont {Hone}, \citenamefont {Dean}, \citenamefont {Zaletel}, \citenamefont {Ashoori},\ and\ \citenamefont {Young}}]{Hunt2017}%
  \BibitemOpen
  \bibfield  {author} {\bibinfo {author} {\bibfnamefont {B.~M.}\ \bibnamefont {Hunt}}, \bibinfo {author} {\bibfnamefont {J.~I.~A.}\ \bibnamefont {Li}}, \bibinfo {author} {\bibfnamefont {A.~A.}\ \bibnamefont {Zibrov}}, \bibinfo {author} {\bibfnamefont {L.}~\bibnamefont {Wang}}, \bibinfo {author} {\bibfnamefont {T.}~\bibnamefont {Taniguchi}}, \bibinfo {author} {\bibfnamefont {K.}~\bibnamefont {Watanabe}}, \bibinfo {author} {\bibfnamefont {J.}~\bibnamefont {Hone}}, \bibinfo {author} {\bibfnamefont {C.~R.}\ \bibnamefont {Dean}}, \bibinfo {author} {\bibfnamefont {M.}~\bibnamefont {Zaletel}}, \bibinfo {author} {\bibfnamefont {R.~C.}\ \bibnamefont {Ashoori}},\ and\ \bibinfo {author} {\bibfnamefont {A.~F.}\ \bibnamefont {Young}},\ }\bibfield  {title} {\bibinfo {title} {Direct measurement of discrete valley and orbital quantum numbers in bilayer graphene},\ }\href {https://doi.org/10.1038/s41467-017-00824-w} {\bibfield  {journal} {\bibinfo  {journal} {Nature Communications}\ }\textbf {\bibinfo {volume} {8}},\ \bibinfo
  {pages} {948} (\bibinfo {year} {2017})}\BibitemShut {NoStop}%
\bibitem [{\citenamefont {Slizovskiy}\ \emph {et~al.}(2021)\citenamefont {Slizovskiy}, \citenamefont {Garcia-Ruiz}, \citenamefont {Berdyugin}, \citenamefont {Xin}, \citenamefont {Taniguchi}, \citenamefont {Watanabe}, \citenamefont {Geim}, \citenamefont {Drummond},\ and\ \citenamefont {Fal'ko}}]{Slizovskiy2021}%
  \BibitemOpen
  \bibfield  {author} {\bibinfo {author} {\bibfnamefont {S.}~\bibnamefont {Slizovskiy}}, \bibinfo {author} {\bibfnamefont {A.}~\bibnamefont {Garcia-Ruiz}}, \bibinfo {author} {\bibfnamefont {A.~I.}\ \bibnamefont {Berdyugin}}, \bibinfo {author} {\bibfnamefont {N.}~\bibnamefont {Xin}}, \bibinfo {author} {\bibfnamefont {T.}~\bibnamefont {Taniguchi}}, \bibinfo {author} {\bibfnamefont {K.}~\bibnamefont {Watanabe}}, \bibinfo {author} {\bibfnamefont {A.~K.}\ \bibnamefont {Geim}}, \bibinfo {author} {\bibfnamefont {N.~D.}\ \bibnamefont {Drummond}},\ and\ \bibinfo {author} {\bibfnamefont {V.~I.}\ \bibnamefont {Fal'ko}},\ }\bibfield  {title} {\bibinfo {title} {Out-of-plane dielectric susceptibility of graphene in twistronic and bernal bilayers},\ }\href {https://doi.org/10.1021/acs.nanolett.1c02211} {\bibfield  {journal} {\bibinfo  {journal} {Nano Letters}\ }\textbf {\bibinfo {volume} {21}},\ \bibinfo {pages} {6678} (\bibinfo {year} {2021})}\BibitemShut {NoStop}%
\bibitem [{\citenamefont {Bessler}\ \emph {et~al.}(2019)\citenamefont {Bessler}, \citenamefont {Duerig},\ and\ \citenamefont {Koren}}]{Bessler2019}%
  \BibitemOpen
  \bibfield  {author} {\bibinfo {author} {\bibfnamefont {R.}~\bibnamefont {Bessler}}, \bibinfo {author} {\bibfnamefont {U.}~\bibnamefont {Duerig}},\ and\ \bibinfo {author} {\bibfnamefont {E.}~\bibnamefont {Koren}},\ }\bibfield  {title} {\bibinfo {title} {The dielectric constant of a bilayer graphene interface},\ }\href {https://doi.org/10.1039/c8na00350e} {\bibfield  {journal} {\bibinfo  {journal} {Nanoscale Advances}\ }\textbf {\bibinfo {volume} {1}},\ \bibinfo {pages} {1702} (\bibinfo {year} {2019})}\BibitemShut {NoStop}%
\bibitem [{\citenamefont {Carr}\ \emph {et~al.}(2018)\citenamefont {Carr}, \citenamefont {Fang}, \citenamefont {Jarillo-Herrero},\ and\ \citenamefont {Kaxiras}}]{Carr2018}%
  \BibitemOpen
  \bibfield  {author} {\bibinfo {author} {\bibfnamefont {S.}~\bibnamefont {Carr}}, \bibinfo {author} {\bibfnamefont {S.}~\bibnamefont {Fang}}, \bibinfo {author} {\bibfnamefont {P.}~\bibnamefont {Jarillo-Herrero}},\ and\ \bibinfo {author} {\bibfnamefont {E.}~\bibnamefont {Kaxiras}},\ }\bibfield  {title} {\bibinfo {title} {Pressure dependence of the magic twist angle in graphene superlattices},\ }\href {https://doi.org/10.1103/physrevb.98.085144} {\bibfield  {journal} {\bibinfo  {journal} {Physical Review B}\ }\textbf {\bibinfo {volume} {98}},\ \bibinfo {pages} {085144} (\bibinfo {year} {2018})}\BibitemShut {NoStop}%
\bibitem [{\citenamefont {Yankowitz}\ \emph {et~al.}(2018)\citenamefont {Yankowitz}, \citenamefont {Jung}, \citenamefont {Laksono}, \citenamefont {Leconte}, \citenamefont {Chittari}, \citenamefont {Watanabe}, \citenamefont {Taniguchi}, \citenamefont {Adam}, \citenamefont {Graf},\ and\ \citenamefont {Dean}}]{Yankowitz2018}%
  \BibitemOpen
  \bibfield  {author} {\bibinfo {author} {\bibfnamefont {M.}~\bibnamefont {Yankowitz}}, \bibinfo {author} {\bibfnamefont {J.}~\bibnamefont {Jung}}, \bibinfo {author} {\bibfnamefont {E.}~\bibnamefont {Laksono}}, \bibinfo {author} {\bibfnamefont {N.}~\bibnamefont {Leconte}}, \bibinfo {author} {\bibfnamefont {B.~L.}\ \bibnamefont {Chittari}}, \bibinfo {author} {\bibfnamefont {K.}~\bibnamefont {Watanabe}}, \bibinfo {author} {\bibfnamefont {T.}~\bibnamefont {Taniguchi}}, \bibinfo {author} {\bibfnamefont {S.}~\bibnamefont {Adam}}, \bibinfo {author} {\bibfnamefont {D.}~\bibnamefont {Graf}},\ and\ \bibinfo {author} {\bibfnamefont {C.~R.}\ \bibnamefont {Dean}},\ }\bibfield  {title} {\bibinfo {title} {Dynamic band-structure tuning of graphene moiré superlattices with pressure},\ }\href {https://doi.org/10.1038/s41586-018-0107-1} {\bibfield  {journal} {\bibinfo  {journal} {Nature}\ }\textbf {\bibinfo {volume} {557}},\ \bibinfo {pages} {404} (\bibinfo {year} {2018})}\BibitemShut {NoStop}%
\bibitem [{\citenamefont {Fülöp}\ \emph {et~al.}(2021{\natexlab{b}})\citenamefont {Fülöp}, \citenamefont {Márffy}, \citenamefont {Zihlmann}, \citenamefont {Gmitra}, \citenamefont {Tóvári}, \citenamefont {Szentpéteri}, \citenamefont {Kedves}, \citenamefont {Watanabe}, \citenamefont {Taniguchi}, \citenamefont {Fabian}, \citenamefont {Schönenberger}, \citenamefont {Makk},\ and\ \citenamefont {Csonka}}]{Fueloep2021}%
  \BibitemOpen
  \bibfield  {author} {\bibinfo {author} {\bibfnamefont {B.}~\bibnamefont {Fülöp}}, \bibinfo {author} {\bibfnamefont {A.}~\bibnamefont {Márffy}}, \bibinfo {author} {\bibfnamefont {S.}~\bibnamefont {Zihlmann}}, \bibinfo {author} {\bibfnamefont {M.}~\bibnamefont {Gmitra}}, \bibinfo {author} {\bibfnamefont {E.}~\bibnamefont {Tóvári}}, \bibinfo {author} {\bibfnamefont {B.}~\bibnamefont {Szentpéteri}}, \bibinfo {author} {\bibfnamefont {M.}~\bibnamefont {Kedves}}, \bibinfo {author} {\bibfnamefont {K.}~\bibnamefont {Watanabe}}, \bibinfo {author} {\bibfnamefont {T.}~\bibnamefont {Taniguchi}}, \bibinfo {author} {\bibfnamefont {J.}~\bibnamefont {Fabian}}, \bibinfo {author} {\bibfnamefont {C.}~\bibnamefont {Schönenberger}}, \bibinfo {author} {\bibfnamefont {P.}~\bibnamefont {Makk}},\ and\ \bibinfo {author} {\bibfnamefont {S.}~\bibnamefont {Csonka}},\ }\bibfield  {title} {\bibinfo {title} {Boosting proximity spin-orbit coupling in graphene/wse2 heterostructures via hydrostatic pressure},\ }\href
  {https://doi.org/10.1038/s41699-021-00262-9} {\bibfield  {journal} {\bibinfo  {journal} {npj 2D Materials and Applications}\ }\textbf {\bibinfo {volume} {5}},\ \bibinfo {pages} {82} (\bibinfo {year} {2021}{\natexlab{b}})}\BibitemShut {NoStop}%
\bibitem [{\citenamefont {P\'eterfalvi}\ \emph {et~al.}(2022)\citenamefont {P\'eterfalvi}, \citenamefont {David}, \citenamefont {Rakyta}, \citenamefont {Burkard},\ and\ \citenamefont {Korm\'anyos}}]{Peterfalvi2022}%
  \BibitemOpen
  \bibfield  {author} {\bibinfo {author} {\bibfnamefont {C.~G.}\ \bibnamefont {P\'eterfalvi}}, \bibinfo {author} {\bibfnamefont {A.}~\bibnamefont {David}}, \bibinfo {author} {\bibfnamefont {P.}~\bibnamefont {Rakyta}}, \bibinfo {author} {\bibfnamefont {G.}~\bibnamefont {Burkard}},\ and\ \bibinfo {author} {\bibfnamefont {A.}~\bibnamefont {Korm\'anyos}},\ }\bibfield  {title} {\bibinfo {title} {Quantum interference tuning of spin-orbit coupling in twisted van der waals trilayers},\ }\href {https://doi.org/10.1103/PhysRevResearch.4.L022049} {\bibfield  {journal} {\bibinfo  {journal} {Phys. Rev. Res.}\ }\textbf {\bibinfo {volume} {4}},\ \bibinfo {pages} {L022049} (\bibinfo {year} {2022})}\BibitemShut {NoStop}%
\bibitem [{\citenamefont {Li}\ and\ \citenamefont {Koshino}(2019)}]{Li2019}%
  \BibitemOpen
  \bibfield  {author} {\bibinfo {author} {\bibfnamefont {Y.}~\bibnamefont {Li}}\ and\ \bibinfo {author} {\bibfnamefont {M.}~\bibnamefont {Koshino}},\ }\bibfield  {title} {\bibinfo {title} {Twist-angle dependence of the proximity spin-orbit coupling in graphene on transition-metal dichalcogenides},\ }\href {https://doi.org/10.1103/PhysRevB.99.075438} {\bibfield  {journal} {\bibinfo  {journal} {Phys. Rev. B}\ }\textbf {\bibinfo {volume} {99}},\ \bibinfo {pages} {075438} (\bibinfo {year} {2019})}\BibitemShut {NoStop}%
\end{thebibliography}%

\end{document}

\author{M\'at\'e Kedves}
\affiliation{Department of Physics, Institute of Physics, Budapest University of Technology and Economics, M\H uegyetem rkp.\ 3., H-1111 Budapest, Hungary}
\affiliation{MTA-BME Correlated van der Waals Structures Momentum Research Group, M\H uegyetem rkp.\ 3., H-1111 Budapest, Hungary}

\author{B\'alint Szentp\'eteri}
\affiliation{Department of Physics, Institute of Physics, Budapest University of Technology and Economics, M\H uegyetem rkp.\ 3., H-1111 Budapest, Hungary}
\affiliation{MTA-BME Correlated van der Waals Structures Momentum Research Group, M\H uegyetem rkp.\ 3., H-1111 Budapest, Hungary}

\author{Albin M\'arffy}
\affiliation{Department of Physics, Institute of Physics, Budapest University of Technology and Economics, M\H uegyetem rkp.\ 3., H-1111 Budapest, Hungary}
\affiliation{MTA-BME Superconducting Nanoelectronics Momentum Research Group, M\H uegyetem rkp.\ 3., H-1111 Budapest, Hungary}

\author{Endre T\'ov\'ari}
\affiliation{Department of Physics, Institute of Physics, Budapest University of Technology and Economics, M\H uegyetem rkp.\ 3., H-1111 Budapest, Hungary}
\affiliation{MTA-BME Correlated van der Waals Structures Momentum Research Group, M\H uegyetem rkp.\ 3., H-1111 Budapest, Hungary}

\author{Nikos Papadopoulos}
\affiliation{QuTech and Kavli Institute of Nanoscience, Delft University of Technology, 2600 GA Delft, The Netherlands}

\author{Prasanna K. Rout}
\affiliation{QuTech and Kavli Institute of Nanoscience, Delft University of Technology, 2600 GA Delft, The Netherlands}

\author{Kenji Watanabe}
\affiliation{Research Center for Functional Materials, National Institute for Materials Science, 1-1 Namiki, Tsukuba 305-0044, Japan}

\author{Takashi Taniguchi}
\affiliation{International Center for Materials Nanoarchitectonics,
National Institute for Materials Science, 1-1 Namiki, Tsukuba 305-0044, Japan}

\author{Srijit Goswami}
\affiliation{QuTech and Kavli Institute of Nanoscience, Delft University of Technology, 2600 GA Delft, The Netherlands}

\author{Szabolcs Csonka}
\affiliation{Department of Physics, Institute of Physics, Budapest University of Technology and Economics, M\H uegyetem rkp.\ 3., H-1111 Budapest, Hungary}
\affiliation{MTA-BME Superconducting Nanoelectronics Momentum Research Group, M\H uegyetem rkp.\ 3., H-1111 Budapest, Hungary}

\author{P\'eter Makk}
\email{makk.peter@ttk.bme.hu}
\affiliation{Department of Physics, Institute of Physics, Budapest University of Technology and Economics, M\H uegyetem rkp.\ 3., H-1111 Budapest, Hungary}
\affiliation{MTA-BME Correlated van der Waals Structures Momentum Research Group, M\H uegyetem rkp.\ 3., H-1111 Budapest, Hungary}

\maketitle

\section{\label{sec:device}Device geometry and measurement setup}

\begin{figure}
\includegraphics[width=0.8\columnwidth]{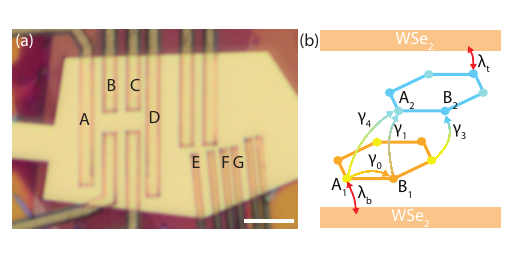}
\caption{\label{fig:device} (a) Optical image of the measure device. The scale bar is 4 $\mu$m. (b) Schematic illustration of the heterostructure, showing the hopping terms in the BLG and the layer dependent SOC terms, induced by the WSe$_2$ layers.}
\end{figure}

The measured sample is shown in Fig.\,\ref{fig:device}. The dry-transfer technique with PC/PDMS hemispheres is employed to stack hBN (35 nm)/WSe$_2$ (19 nm)/BLG/WSe$_2$ (19 nm)/hBN (60 nm)/graphite. To fabricate electrical contacts to the Hall bar, we use e-beam lithography patterning followed by a reactive ion etching step using CHF$_3$/O$_2$ mixture and finally deposit Ti (5nm)/NbTiN (100 nm) by dc sputtering. We deposit Al$_2$O$_3$ (30 nm) using ALD which acts as the gate dielectric. Finally, the top gate is defined by e-beam lithography and deposition of Ti (5 nm)/Au (100 nm). We note that the heterostructure was not etched into a Hall-bar shape after the contacts were deposited. Therefore current can flow along the pristine edges of the BLG layer. For the quantum Hall measurements, this results in a nontrivial sample geometry which could result in the mixing of longitudinal and transverse resistances. During the fabrication process, the alignment of WSe$_2$ layers was not controlled.

Transport measurements were carried out in an Oxford cryostat equipped with a variable temperature insert (VTI) at a base temperature of 1.4 K (unless otherwise stated). Measurements were performed using lock-in technique with 0.1 mV AC voltage excitation at 1171 Hz frequency. Measurements presented in the main text were conducted on device S1: the AC voltage bias was applied between contacts A and D while the four-terminal voltages were measured between B and C.

To apply hydrostatic pressure, the sample is first bonded to a high pressure sample holder and placed in a piston-cylinder pressure cell as detailed in Ref.\,\cite{Fueloep2021a} where kerosene acts as the pressure mediating medium. To change the applied pressure the sample is warmed up to room temperature where the pressure is applied using a hydraulic press and the sample is cooled down
again.

\section{\label{sec:theory}Continuum model of WSe2/BLG/WSe2}
\subsection{Low-energy Hamiltonian}

We model the studied heterostructure with the low-energy Hamiltonian of bilayer graphene with an additional spin--orbit coupling term, which is different on the two graphene layers, induced by the proximity of the two WSe$_2$ layers. In the basis of the 4 atom unit cell, 
$\left(\ket{C_\mathrm{A1}}, \ket{C_\mathrm{B1}}, \ket{C_\mathrm{A2}}, \ket{C_\mathrm{B2}}\right)\otimes \left(\ket{\uparrow},\ket{\downarrow}\right)$ the Hamiltonian is written as 
\begin{eqnarray}
\mathcal{H}&=&\mathcal{H}_\mathrm{BLG}+\mathcal{H}_\mathrm{SOC},\\
\mathcal{H}_\mathrm{BLG}&=&\left(\begin{array}{cccc}
u/2&v_0\pi^\dagger&-v_4\pi^\dagger&v_3\pi\\
v_0\pi&u/2+\Delta' &\gamma_1&-v_4\pi^\dagger\\
-v_4\pi&\gamma_1&-u/2+\Delta'&v_0\pi^\dagger\\
v_3\pi^\dagger&-v_4\pi&v_0\pi&-u/2
\end{array}
\right)\otimes s_0,
\label{H_BLG}\\
\mathcal{H}_\mathrm{SOC}&=&\left(\begin{array}{cccc}
\xi\lambda_I^{b} s_z/2 & i\lambda_R^{b} s_-^\xi&0&0\\
-i\lambda_R^{b} s_+^\xi&\xi\lambda_I^{b} s_z /2&0&0\\
0&0&\xi\lambda_I^{t} s_z/2 & i\lambda_R^{t} s_-^\xi\\
0&0&-i\lambda_R^{t} s_+^\xi&\xi\lambda_I^{t} s_z /2
\end{array}
\right),
\end{eqnarray}
where $\mathcal{H}_\mathrm{BLG}$ is the Hamiltonian of the BLG\cite{McCann2013} and $\mathcal{H}_\mathrm{SOC}$ is the spin--orbit coupling term describing the proximity induced Ising-type SOC with the parameters of $\lambda_I^i$ and Rashba-type SOC parametrized with $\lambda_R^i$ \cite{Konschuh2012,Zollner2021}. Here, $s_i$, with $i=\{0,x,y,z\}$, are the spin Pauli matrices and $s_\pm^\xi=\frac{1}{2}(s_x+i\xi s_y)$. In $\mathcal{H}_\mathrm{BLG}$, $\gamma_i$, with $i=\{0,1,3,4\}$ describe the intra- and interlayer hoppings in BLG, as illustrated in Fig.~\ref{fig:device}.b, $v_i=\sqrt{3}a\gamma_i/2\hbar$ are effective velocities, with the lattice constant of the graphene $a=2.46$\,\AA\ and $\Delta'$ is the dimer on-site energy. $\gamma_0$ is the nearest neighbour intralayer hopping, $\gamma_1$ is the interlayer hopping between the dimer sites, $\gamma_3$ describes the hopping between the non-dimer sites and $\gamma_4$ is the interlayer hopping between the dimer and non-dimer orbitals. In 
$\mathcal{H}$, $\pi=\hbar(\xi k_x+ik_y)$ and $\pi^\dagger=\hbar(\xi k_x-ik_y)$ are momentum operators measured from the K and K' valleys with the valley indices $\xi=\pm1$. The parameter $u$ is the interlayer potentials difference modelling the effect of an external electric field. 

In our simulations we have used the following parameters: $\gamma_0=2.61$\,eV, $\gamma_1=0.361$\,eV, $\gamma_3=0.283$\,eV, $\gamma_4=0.138$\,eV and $\Delta'=0.015$\,eV \cite{Jung2014}.

\begin{figure}[!hbt]\centering
\includegraphics[width=0.8\columnwidth]{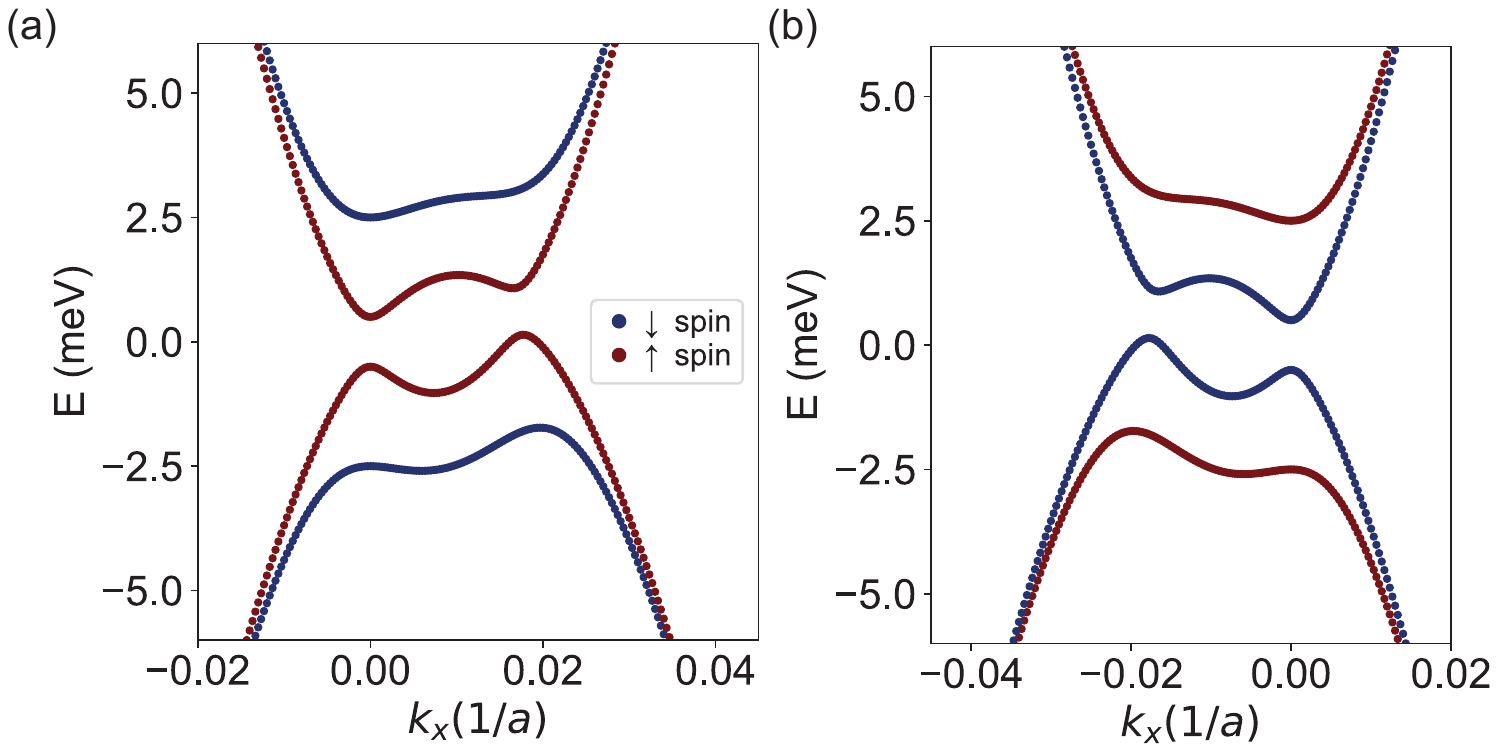}
\caption{{Calculated band structure using the parameters of $u=3$\,meV and $\lambda_I^{b}=-\lambda_I^{t}=2$\,meV (a) near the $K$ valley and (b) near the $K'$ valley. }}
\label{supfig:K_Kp_compare}
\end{figure}

In the main text, we show the spectrum near the $K$ valley. Here, we show the difference between the $K$ and $K'$ valleys in Fig~\ref{supfig:K_Kp_compare}. The main difference, besides the opposite tilting due to the trigonal warping, is the opposite spin polarization of the bands, which is the manifestation of the Kramers theorem, since the valley-Zeemann terms generate an opposite effective magnetic field in the two valleys. In the figures, we calculate the spin polarization as
\begin{equation}
\zeta_n=\sum\limits_{X=A1,A2,B1,B2}|c_{X,\uparrow}|^2-|c_{X,\downarrow}|^2.
\end{equation}

\begin{figure}[!hbt]\centering
\includegraphics[width=0.8\columnwidth]{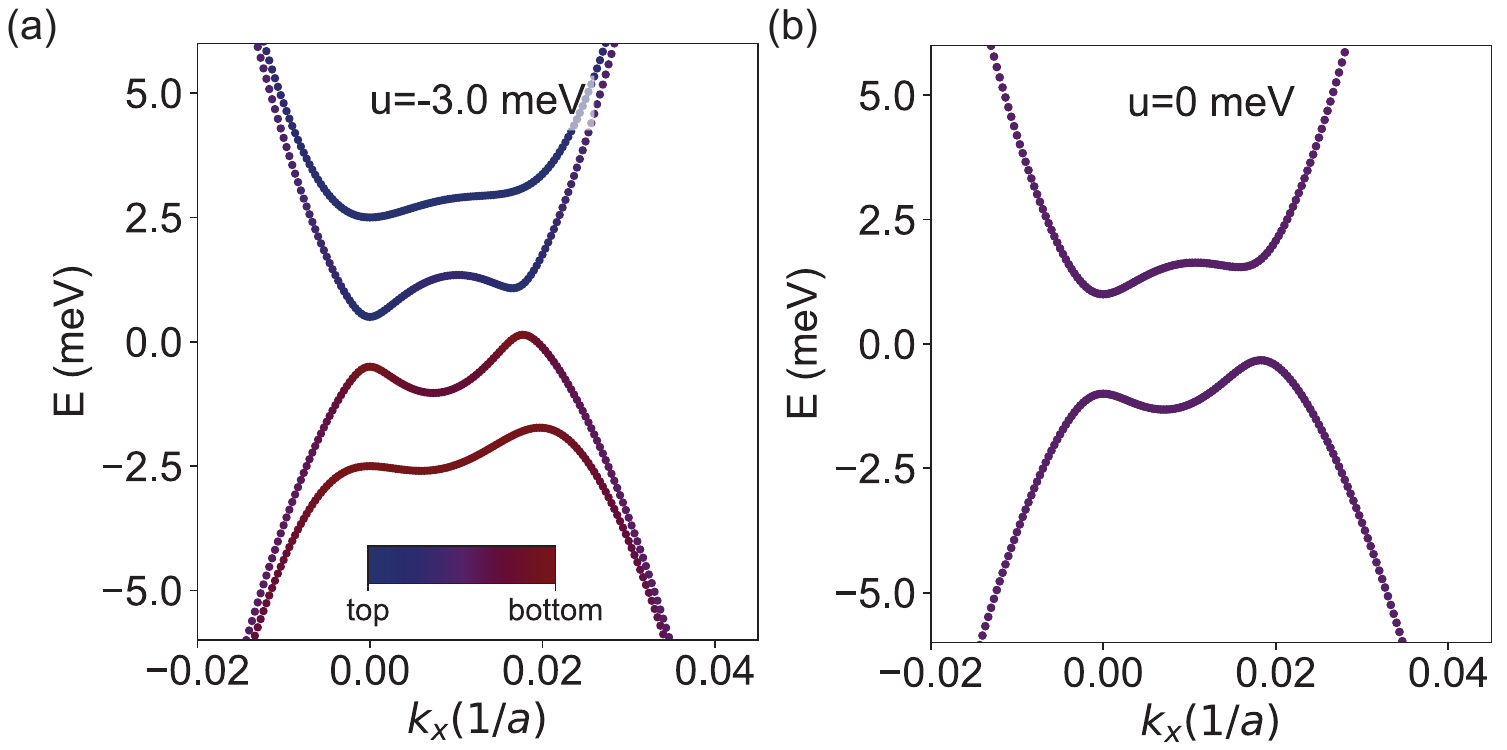}
\caption{{Calculated band structure using the parameter of $\lambda_I^{b}=-\lambda_I^{t}=2$\,meV (a) at $u=-3$\,meV  and (b) at $u=0$\,meV. The color of the line corresponds to the layer polarization: the blue (red) points are fully polarized to the top (bottom) layer and the purple points are layer degenerated.}}
\label{supfig:layerpol}
\end{figure}

Besides the spin polarization, the layer polarization $\alpha_n$ is also an important parameter of the model, which is defined as
\begin{equation}
\alpha_n=\sum\limits_{s=\uparrow,\downarrow}|c_{A1,s}|^2+|c_{B1,s}|^2-|c_{A2,s}|^2-|c_{B2,s}|^2.
\end{equation}
As shown in Fig.~\ref{supfig:layerpol}. at $u=0$ the bands have no layer polarization, which can be lifted by increasing $u$. For $|u|>|\lambda_I^{b}|$, the low energy part of the conductance and valence bands becomes layer polarized with the opposite layer polarization of the valence and conduction bands.

\begin{figure}[!hbt]\centering
\includegraphics[width=0.8\columnwidth]{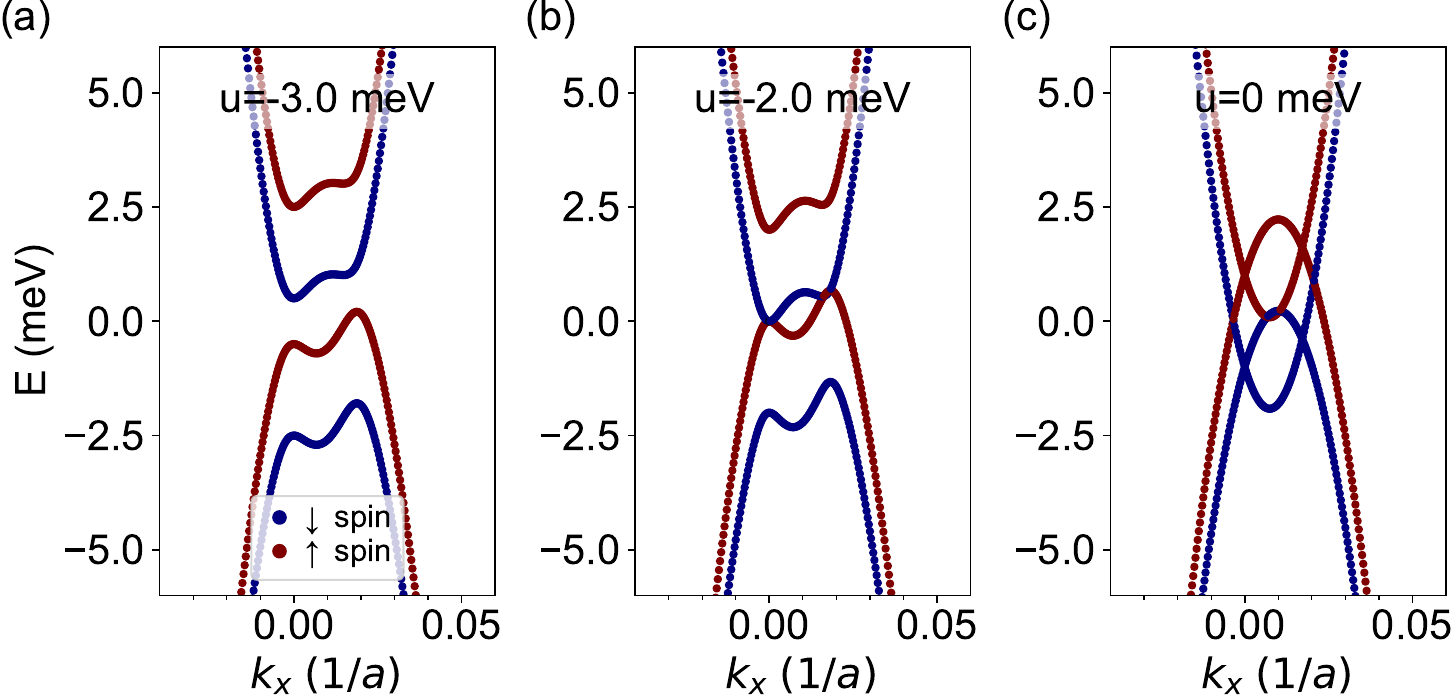}
\caption{{Calculated band structure using the parameter of $\lambda_I^{b}=\lambda_I^{t}=2$\,meV (a) at $u=-3$\,meV, (b) at $u=-2$\,meV  and (c) at $u=0$\,meV. The color of the lines corresponds to the spin polarization.}}
\label{supfig:sameSOC}
\end{figure}

For completeness, the band structure near the $K$-point is shown in Fig.~\ref{supfig:sameSOC}. with $\lambda_I^{b}=\lambda_I^{t}$. In this case, the bands are spin split due to the SOC, which can be considered here as a Zeemann splitting in an effective magnetic field. Moreover, in this case, there is no gap at $u=0$ and a gap only opens if $|u|>|\lambda_I^{b,t}|$. 

\subsection{Landau level calculations}
In high perpendicular magnetic field $B$, we replace the canonical impulse with the kinetic momentum $\hbar q_i=\hbar k_i-eA_i$, where $A_i$ is the vector potential and introduce the magnetic ladder operators as $\hat{a}=\sqrt{\frac{\hbar}{2eB}}\left(q_x+iq_y\right)$ and $\hat{a}^\dagger=\sqrt{\frac{\hbar}{2eB}}\left(q_x-iq_y\right)$, which satisfy $[\hat{a},\hat{a}^\dagger]=1$. In Eq. \ref{H_BLG}, we replace $\pi$ and $\pi^\dagger$ with the ladder operators as $\pi=\sqrt{2eB\hbar}\hat{a}$ and $\pi^\dagger=\sqrt{2eB\hbar}\hat{a}^\dagger$ for the $K$ valley and $\pi=-\sqrt{2eB\hbar}\hat{a}^\dagger$ and $\pi^\dagger=-\sqrt{2eB\hbar}\hat{a}$ for the $K'$ valley. We neglect $\gamma_3$ in the Landau level (LL) calculations, which would introduce a small mixing of the zero LLs and the higher LLs. For the $K$ valley the $\mathcal{H}_\mathrm{BLG}$ can be rewritten as
\begin{equation}
\mathcal{H}_\mathrm{BLG}^K=\left(\begin{array}{cccc}
u&v_0\sqrt{2eB\hbar}\hat{a}^\dagger&-v_4\sqrt{2eB\hbar}\hat{a}^\dagger&0\\
v_0\sqrt{2eB\hbar}\hat{a}&u+\Delta' &\gamma_1&-v_4\sqrt{2eB\hbar}\hat{a}^\dagger\\
-v_4\sqrt{2eB\hbar}\hat{a}&\gamma_1&-u+\Delta'&v_0\sqrt{2eB\hbar}\hat{a}^\dagger\\
0&-v_4\sqrt{2eB\hbar}\hat{a}&v_0\sqrt{2eB\hbar}\hat{a}&-u
\end{array}\right)\otimes s_0,
\end{equation}
and for $K'$ valley it is given by
\begin{equation}
\mathcal{H}_\mathrm{BLG}^{K'}=\left(\begin{array}{cccc}
u&-v_0\sqrt{2eB\hbar}\hat{a}&v_4\sqrt{2eB\hbar}\hat{a}&0\\
-v_0\sqrt{2eB\hbar}\hat{a^\dagger}&u+\Delta' &\gamma_1&v_4\sqrt{2eB\hbar}\hat{a}\\
v_4\sqrt{2eB\hbar}\hat{a}^\dagger&\gamma_1&-u+\Delta'&-v_0\sqrt{2eB\hbar}\hat{a}\\
0&v_4\sqrt{2eB\hbar}\hat{a}^\dagger&-v_0\sqrt{2eB\hbar}\hat{a}^\dagger&-u
\end{array}\right)\otimes s_0.
\end{equation}

The full Hamiltonian in $B$ field is also expanded with the Zeeman term $\mathcal{H}_Z=E_Zs_z$, where $E_Z=-\mu_BB$ is the Zeeman energy with the Bohr magneton $\mu_B$. The eigenenergies $E_{\xi,n,s_z}$ of  $\mathcal{H}=\mathcal{H}_\mathrm{BLG}+\mathcal{H}_\mathrm{SOC}+\mathcal{H}_Z $ are defined as
\begin{equation}
\label{eq:LLeig}
\mathcal{H}\ket{\xi,n,\sigma}=E_{\xi,n,s_z}\ket{\xi,n,\sigma},
\end{equation}
with the eigenstates of $\ket{\xi,n,\sigma}$. The ladder operators act on the Landau level wavefunctions as $\hat{a}\ket{n}=\sqrt{n}\ket{n-1}$ and $\hat{a}^\dagger\ket{n+1}=\sqrt{n+1}\ket{n+1}$. Following the footsteps of \citep{Khoo2018}, the same ansatz can be used to solve Eq. \ref{eq:LLeig}, which was used with induced SOC only in one graphene layer.

\begin{figure}[!hbt]\centering
\includegraphics[width=1\columnwidth]{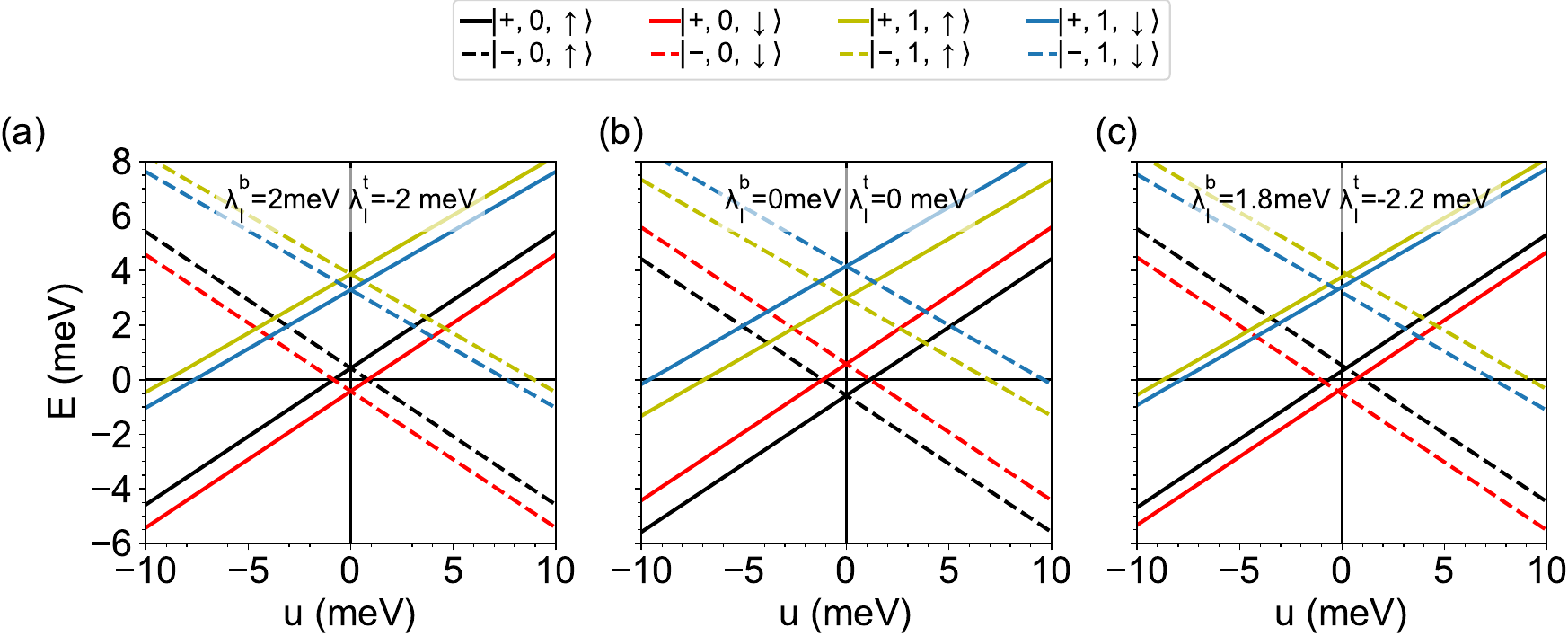}
\caption{{Single-particle zeroth LL spectrum as a function of $u$ at $B=10$\,T for (a) $\lambda_I^b=-\lambda_I^t=-2$\,meV, (b) $\lambda_I^b=\lambda_I^t=-0$ and (c) $\lambda_I^b=1.8$\,meV and $\lambda_I^t=-2.2$\,meV.}}
\label{supfig:LLspectrum}
\end{figure}
By solving the eigenvalue problem, the single-particle Landau level energies can be obtained. In Fig.~4a. in the main text, we plot the lowest 8 LLs with respect to $u$ at $B=8.5$\,T. These LLs would be degenerate if we set every parameter to zero except $\gamma_0$ and $\gamma_1$, including the interlayer potential. In Fig.~\ref{supfig:LLspectrum} we show the LLs in three different scenarios of $\lambda_I$: in panel (b) the LLs are shown if $\lambda_I=0$. In this case, the LLs are spin split due to the Zeeman term and also split in the orbital index due to $\gamma_0$ and $\Delta'$. A finite $u$ further splits these LLs and their energy is linear in $u$. Since for the two lowest LL the spin and layer index becomes effectively the same, different valleys shift oppositely with a displacement field. By introducing a finite Ising-type SOC $\lambda_I^b$ shifts the energies of the eigenstates of $\ket{K,n,\sigma}$ and $\lambda_I^t$ shifts the energies of the eigenstates of $\ket{K',n,\sigma}$. Comparing panel (a) and (b), if $\lambda_I^t=-\lambda_I^b$, at high magnetic field, the spectrum seems very similar to the case without SOC, however, the order of the spin up and spin down levels flip. Moreover, the positions ($u^*$) where two LLs cross also change, which we defined as $D^*$ crossings in the measurements in the main text. In the third case, shown in panel (c), when $\lambda_I^t$ and $\lambda_I^b$ have an opposite sign but their magnitudes are different, the $u$, $-u$ symmetry is lost leading to a non-zero $\pm u_3^*$ crossings.

\begin{figure}[!hbt]\centering
\includegraphics[width=\columnwidth]{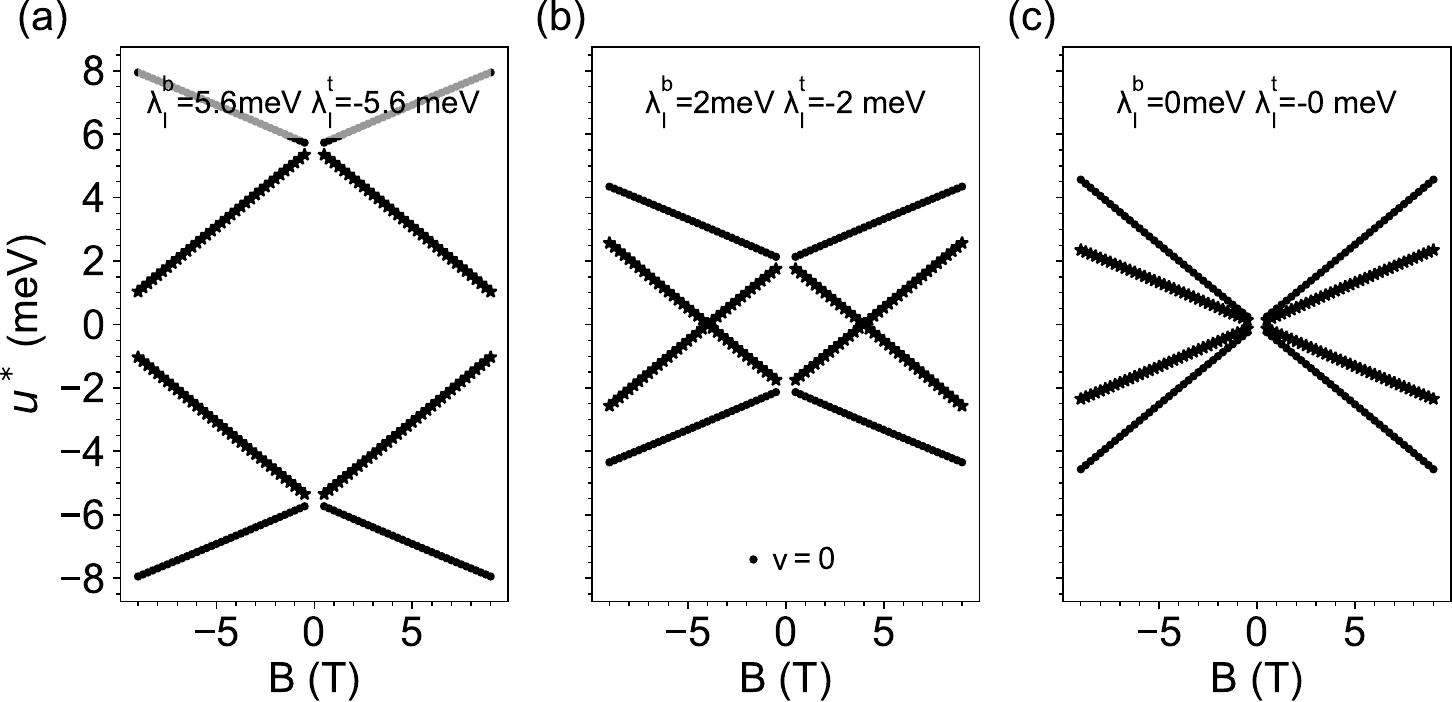}
\caption{Landau level crossings at $\nu=0$ as a function of $B$ with (a) $\lambda_I^b=-\lambda_I^t=-5.6$\,meV, (b) $\lambda_I^b=-\lambda_I^t=-2$\,meV and (c) $\lambda_I^b=\lambda_I^t=-0$.}
\label{supfig:LLcrossing0}
\end{figure}
\begin{figure}[!hbt]\centering
\includegraphics[width=\columnwidth]{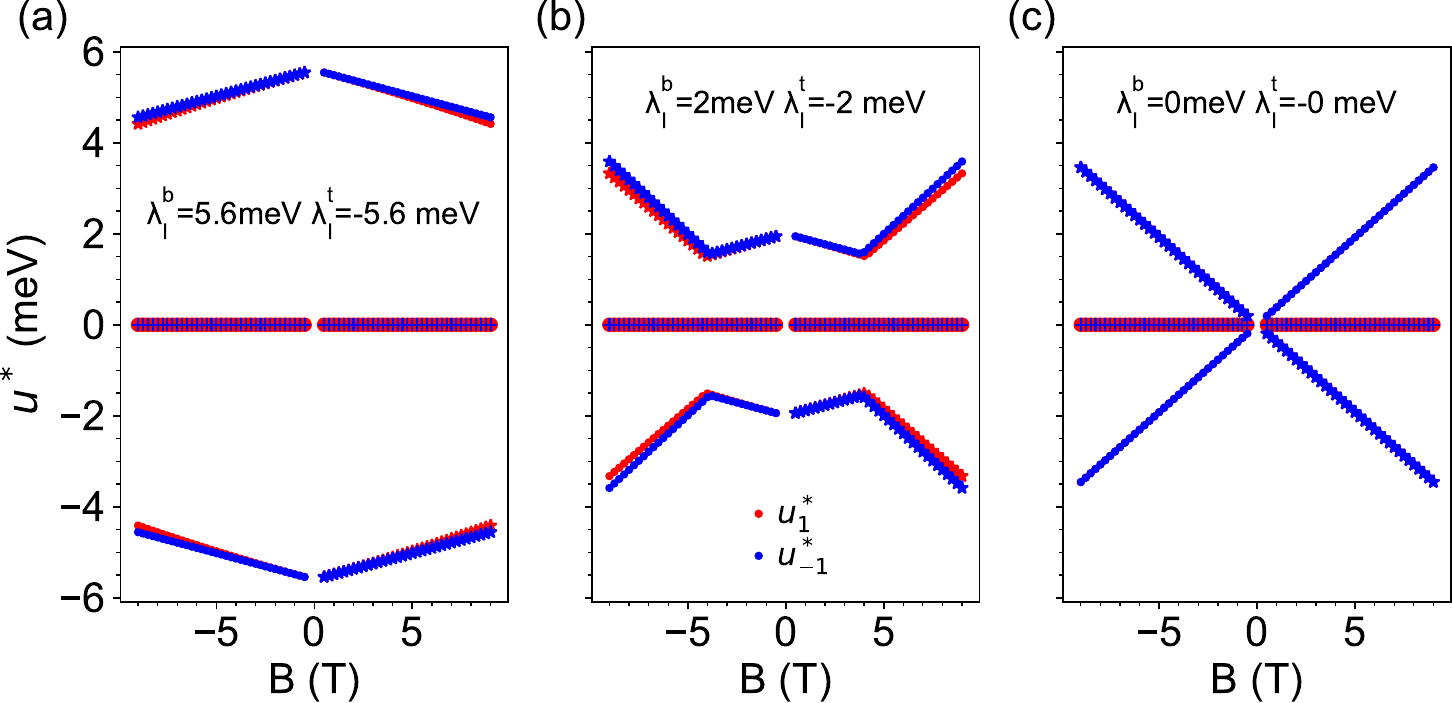}
\caption{Landau level crossings at $\nu=\pm1$ as a function of $B$ with (a) $\lambda_I^b=-\lambda_I^t=-5.6$\,meV, (b) $\lambda_I^b=-\lambda_I^t=-2$\,meV and with (c) $\lambda_I^b=\lambda_I^t=0$.}
\label{supfig:LLcrossing1}
\end{figure}

In Fig.~\ref{supfig:LLcrossing0}. and Fig.~\ref{supfig:LLcrossing1}. we plot the Landau level crossings, with SOC and without SOC, for $\nu=0$ and $\nu=\pm1$, respectively. Without SOC the crossings go to zero as $B\to0$ as opposed to the case of $\lambda_I^b=-\lambda_I^t\neq0$. Comparing these figures with Fig~4. in the main text, the experiments show similarities to the model: the $\nu=1$ crossings show similar tendency, so do the higher $u^*_0$ branches in the $\nu=0$ crossings. Likely the discrepancy between the model calculated in a single-particle picture and the experiment comes from the fact that in our calculations we neglect electron-electron interactions \citep{Hunt2017}.

\section{\label{sec:level_arms}Determination of the lever arms}
\begin{figure}
\includegraphics[width=\columnwidth]{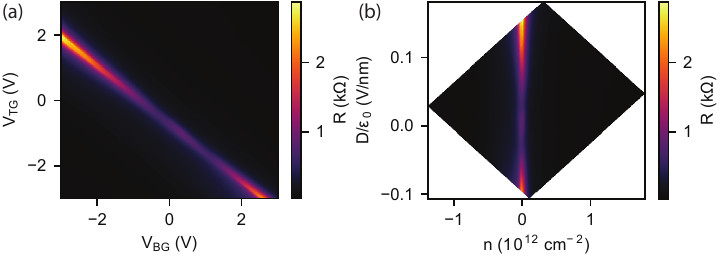}
\caption{\label{fig:nD_convert} Conversion from a) gate voltages to b) charge carrier density $n$ and transverse displacement field $D$.}
\end{figure}

As discussed in the main text, to tune the charge carrier density $n$ and the transverse displacement field $D$ in the sample, gate voltages are applied to the metallic topgate ($V_{TG}$) and the graphite bottom gate ($V_{BG}$). The conversion from gate voltages to $n$ and $D$ is shown in Fig.\,\ref{fig:nD_convert} and is given by the following relation:

\begin{equation}
    \begin{split}
        n & = \alpha_{TG}V_{TG}+\alpha_{BG}V_{BG}+n_0 \\
        \frac{D}{\epsilon_0} & =\frac{e}{2\epsilon_0}\left(\alpha_{TG}V_{TG}-\alpha_{BG}V_{BG}\right)+\frac{D_0}{\epsilon_0},
    \end{split}
\end{equation}
where $\epsilon_0$ is the vacuum permittivity, $e$ is the elementary charge, $\alpha_{BG,TG}$ are the lever arms of the bottom and topgate, respectively, while $n_{0}$ and $D_{0}$ are the offset charge carrier density and displacement field. Since the lever arms are subject to change after the hydrostatic pressure is applied, originating from the compression of dielectricts, we determine them experimentally. First, the ratio of lever arms $\alpha_{BG}/\alpha_{TG}$ can be obatined from gate voltage maps of the resistance (e.g. Fig.\,\ref{fig:nD_convert}.a), as it is given by the slope of the charge neutrality line. Secondly, by measuring the fan diagram of Landau levels for $D=0$, we determine the lever arms via the relation $\nu=nh/eB$ between the filling factor $\nu$ and the carrier density $n$ for a given magnetic field $B$, where $h$ is Planck's constant. The values of the lever arms can be found in Table\,\ref{tab:lever_arms}.

\begin{figure}
\includegraphics[width=\columnwidth]{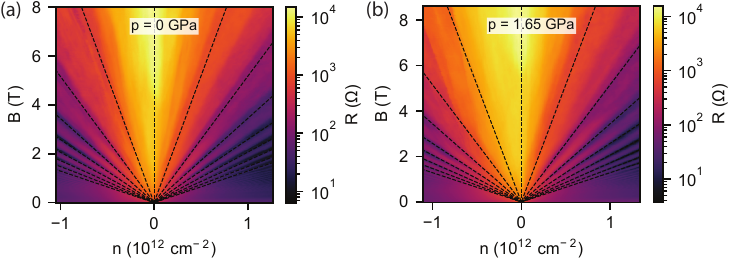}
\caption{\label{fig:fan_plot} Landau fan daigram of the resistance for a) $p=0$ and b) $p=1.65$\,GPa at $D=0$. Dashed lines correspond to carrier densities with filling factors $\nu=4k$, where $k\in\mathbb{Z}$.}
\end{figure}

\begin{table}
\centering
\begin{tabular}{|c |c |c |} 
 \hline
 $p$ ($GPa$) & $\alpha_{BG}$ ($10^{15}$ V$^{-1}$m$^{-2}$)& $\alpha_{TG}$  ($10^{15}$ V$^{-1}$m$^{-2}$)\\
 \hline
 0 & $2.47\pm0.08$ & $2.81\pm0.10$ \\ 
 \hline
 1.65 & $2.82\pm0.07$ & $3.12\pm0.09$ \\
\hline
\end{tabular}

\caption{\label{tab:lever_arms} Lever arms determined from quantum Hall measurements.}
\end{table}

\section{\label{sec:Activation}Extended activation data}
\begin{figure}
\includegraphics[width=0.7\columnwidth]{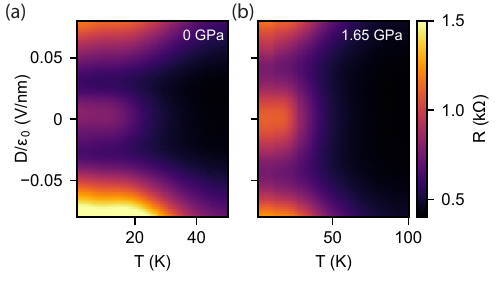}
\caption{\label{fig:activation_data} Temperature dependence of the resistance $R$ as a function of $D$ at $n=0$ for a) $p=0$ and b) $p=1.65$ GPa.}
\end{figure}

Raw measurement data obtained while cooling the device is shown in Fig.\,\ref{fig:activation_data} for $p=0$ and $p=1.65$\,GPa. Similar maps were recorded while warming up the device from base temperature (not shown).

\section{\label{sec:SOI_strength}Determination of the SOI parameters}
\begin{figure}
\includegraphics[width=0.75\columnwidth]{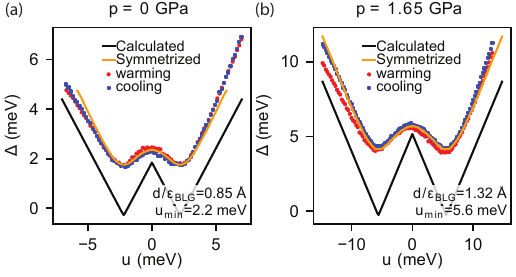}
\caption{\label{fig:activation} Band gaps determined from thermal activation measurements performed while warming up (red) and cooling down (blue) the device for a) $p=0$ and b) $p=1.65$\,GPa, respectively. Symmetrized curve with respect to $u=0$ is shown in orange (see text for details) and the band gap calculated from the theoretical model is shown with the solid black line.}
\end{figure}

To obtain the SOI strength from thermal activation measurements, electric displacement field $D$ has to be converted to the induced interlayer potential difference $u$. As described in the main text, this is done via the relation $u=-\frac{ed}{\epsilon_0 \epsilon_\mathrm{BLG}}D$. For ambient pressure, $d=3.3$\,\r{A} can be taken, similarly to pristine BLG.  On the other hand, the value of $\epsilon_\mathrm{BLG}$ available in the literature ranges from 2.6\,\cite{Slizovskiy2021} to 6\,\cite{Bessler2019}. Since in the large $u$ limit, the band gap induced by the displacement field is independent of the SOI parameters, by varying $\epsilon_\mathrm{BLG}$, we can effectively "fit" our model to the experimental data. Fig.\,\ref{fig:activation}.a shows the extended thermal activation data (partially presented in the main text) measured while warming up (red symbols) and cooling down (blue symbols) the sample for $p=0$. As it is visible, in the band insulator regimes for $u\ll 0$ and $u\gg 0$, the data have different slopes. We take this effect into account by averaging of the two measurements and symmetrizing it with respect to $u=0$ (solid orange line). In the next step, we determine $\epsilon_\mathrm{BLG}$ by matching the slope of the high-$u$ part of the data to match the slope of the theoretical model (solid black line). Finally, we numerically determine the position $u_{min}$ of the band gap minimum of the symmetrized curve and take the Ising SOI parameters as $\lambda_t=-\lambda_b=u_{min}$. We estimate the lower and upper bound of the SOI parameter by fitting the slope of the theoretical curve to the $u\ll 0$ and $u\gg 0$ parts of averaged band gaps, respectively, yielding $\lambda_t=-\lambda_b=2.2\pm0.4$\,meV for $p=0$.

\begin{figure}
\includegraphics[width=0.75\columnwidth]{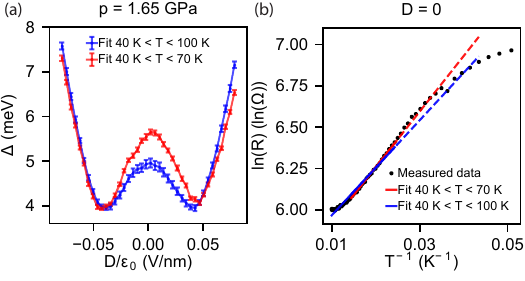}
\caption{\label{fig:activation_fit} a) Band gaps determined from thermal activation measurements performed while cooling down dev S1 at $p=1.65$\,GPa, using a linear fit to the data according to the Arrhenius formula ($\ln R \propto \Delta/2k_B T$) in the temperature range 40\,K$<T<$\,70\,K (blue) and 40\,K$<T<$\,100\,K (red). Error bars represent the uncertainty of the fits. b) Examples for linear fits at $D=0$ in the temperature range 40\,K$<T<$\,70\,K (blue) and 40\,K$<T<$\,100\,K (red) are shown with solid lines and are extended with dashed lines for better visibility. In this case, the difference in the slopes stems from the saturation of the resistance around 100 K.}
\end{figure}

In contrast to ambient pressure, at $p=1.65$\,GPa, we expect both $d$ and $\epsilon_{\mathrm{BLG}}$ to change due to applied pressure. From theoretical predictions\,\cite{Carr2018,Yankowitz2018}, we expect a change in the BLG interlayer distance of $\Delta d<5\%/$GPa. However, to estimate the change in $\epsilon_{\mathrm{BLG}}$ is more challenging. To be able to extract the SOI strength at $p=1.65$\,GPa, we vary the $d/\epsilon_\mathrm{BLG}$ ratio to match the experimental data to the model using the method described above. This way, for $p=1.65$\,GPa, we extract $\lambda_t=-\lambda_b=5.6\pm0.6$\,meV. The extracted increase in SOI strength is comparable to theoretical predictions using ab initio calculations\cite{Fueloep2021}. Furthermore, the SOI parameters extracted from the minima give the same order of magnitude estimate as the extracted gaps at $u=0$. Therefore it is clear, that the relative increase of the positions of band gap minima in $D$ cannot alone stem from changes in the $d/\epsilon_\mathrm{BLG}$ ratio.

To estimate the robustness of our method, we extracted the band gaps from the thermal activation measurements at $p=1.65$\,GPa using linear fits to different temperature ranges. These are shown in Fig.\,\ref{fig:activation_fit}. for ranges of 40\,K$<T<$\,70\,K (blue) and 40\,K$<T<$\,100\,K. The error bars on the figure show the error of the fit for given $u$ and fixed temperature range. From this, we conclude that the uncertainty of the extracted band gaps is $<20\%$. More importantly, the uncertainty of the slope of the $u\ll 0$ and $u\gg 0$ regimes and the positions of band gap minima is even less. Since in our analysis, these are the parameters that determine the extracted SOI strength, this results in an uncertainty of the $d/\epsilon_\mathrm{BLG}$ ratio and the SOI strength of $\sim10\%$ that is comparable to the uncertainty estimated from the difference in the slopes of the $u\ll 0$ and $u\gg 0$ regimes.

\begin{figure}
\includegraphics[width=1\columnwidth]{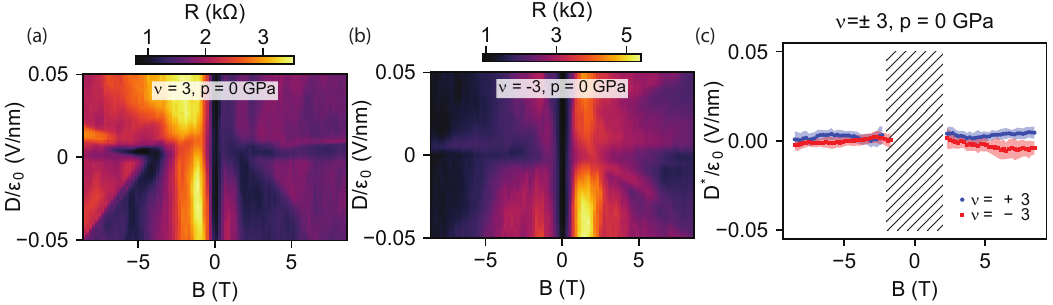}
\caption{\label{fig:nu3} a,b) Measurements of LL crossings as a function of $B$ for $\nu = +3$ and $\nu = -3$, respectively, for
$p = 0$. c) Positions of the LL crossings extracted from the maps in a) and b) for $\nu=+3$ (blue) and $\nu=-3$ (red).}
\end{figure}

We also have to note that, although we assumed that $|\lambda_{I}^{b}|=|\lambda_{I}^{t}|$, our method of determining the SOI parameters is only sensitive to the absolute difference of the two parameters since this quantity defines the closing of the band gap. In other words, the minima of the $\Delta(u)$ functions shown in Fig.\,\ref{fig:activation} are insensitive to a difference in the abolute values of $|\lambda_{I}^{b,t}|$ as long as $\left|\lambda_{I}^{t}-\lambda_{I}^{b}\right|/2$ is constant. However, we can estimate the asymmetry of $|\lambda_{I}^{b,t}|$ by measuring the $\nu=\pm3$ LL crossings since, within our model, the positions of these crossings in $u$ are separated by $u^{*}_{+3}-u^{*}_{-3}\approx|\lambda_{I}^{t}|-|\lambda_{I}^{b}|$, nearly insensitive to the magnetic field. Measurements of the LL crossings as a function of $D$ and $B$ for $\nu=+3$ and $\nu=-3$ are shown in Fig.\,\ref{fig:nu3}.a,b, respectively. Fig.\,\ref{fig:nu3}.c shows the extracted positions $D^{*}$ of the crossings for $\nu=\pm3$. As it is visible, for the most part of our magnetic field range, the positions of the crossings are indistinguishable. At higher magnetic field we see crossings at $u^{*}_{\pm3}\neq0$, indicating a small asymmetry of the SOC parameters. From these, we can estimate the upper bound of the asymmetry as $|\lambda_{I}^{t}|-|\lambda_{I}^{b}|<0.4$\,meV for $p=0$. 

\section{Extended quantum Hall measurement data}
\begin{figure}
\includegraphics[width=1\columnwidth]{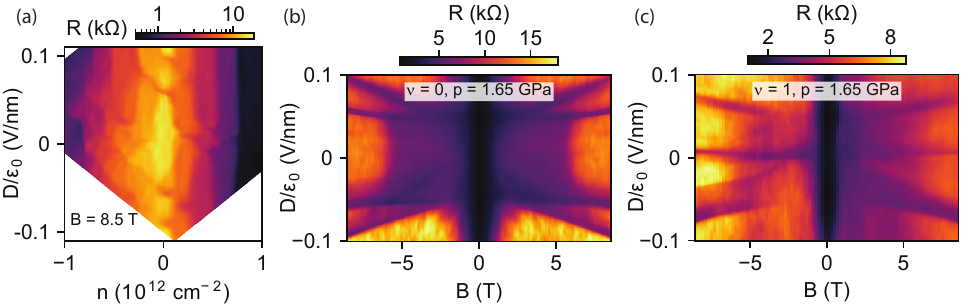}
\caption{\label{fig:crossingsatpress} Extended data for Fig.\,4. of the main text. a) $n$-$D$ map of the resistance measured at $B=8.5$\,T and $p=1.65$\,GPa. b,c) Measurements of LL crossings as a function of $B$ for $\nu = 0$ and $\nu = 1$, respectively, at
$p = 1.65$\,GPa.}
\end{figure}

Additional data for the quantum Hall measurements at $p=1.65$\,GPa is shown in Fig.\,\ref{fig:crossingsatpress}.

\section{Additional measurements on two-terminal devices}
\begin{figure}
\includegraphics[width=1\columnwidth]{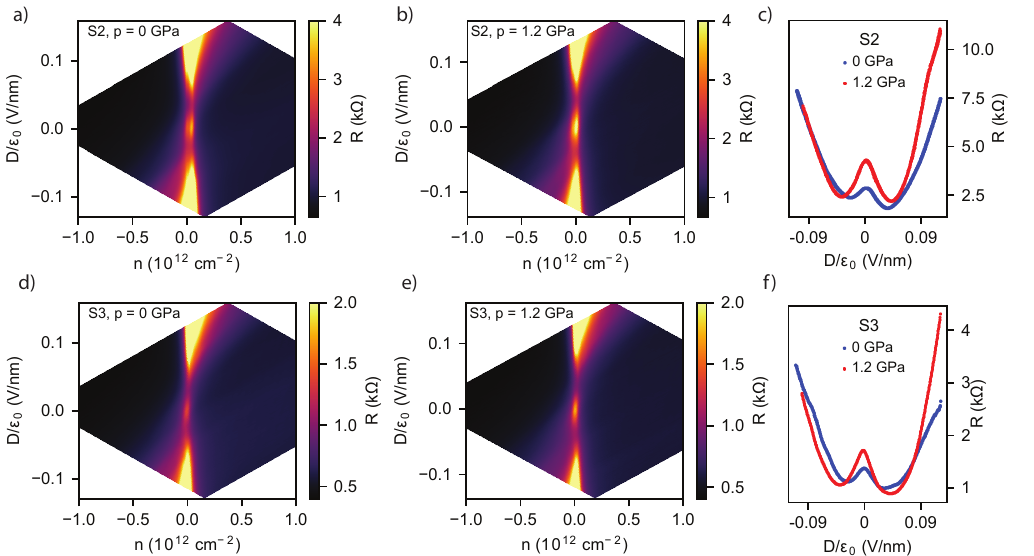}
\caption{\label{fig:gatemaps_S2_S3} a,b) 2-terminal resistance map of dev S2 for a) ambient pressure and b) $p=1.2$\,GPa. c) Line traces of resistance for dev S2 as a function of $D$ at $n=0$ for ambient pressure (blue) and $p=1.2$\,GPa (red). d,e,f) Similar resistance maps and comparison for dev S3.}
\end{figure}

We performed additional measurements on two-terminal devices also shown in Fig.\,\ref{fig:device}.a. Two different devices were measured between contacts E-F (dev S2) and F-G (dev S3) in separate pressurization and cool-down cycles. Two-terminal resistance maps of the devices as a function of $n$ and $D$ are shown in Fig.\,\ref{fig:gatemaps_S2_S3} at $p=0$ and $p=1.2$\,GPa. These results are consistent with our observations on dev S1 in that the inverted phase is clearly visible and is enhanced by the applied pressure. Fig.\,\ref{fig:gatemaps_S2_S3}.c and \ref{fig:gatemaps_S2_S3}.f shows the comparison between line traces of the resistance measured at $n=0$ as a function of $D$ at $p=0$ and $p=1.2$\,GPa for dev S2 and S3, respectively. In both cases, the resistance at $D=0$ is increased due to the applied pressure and the location of resistance minima ($D^{*}$) is increased by $\sim 25\%$.

\begin{figure}
\includegraphics[width=1\columnwidth]{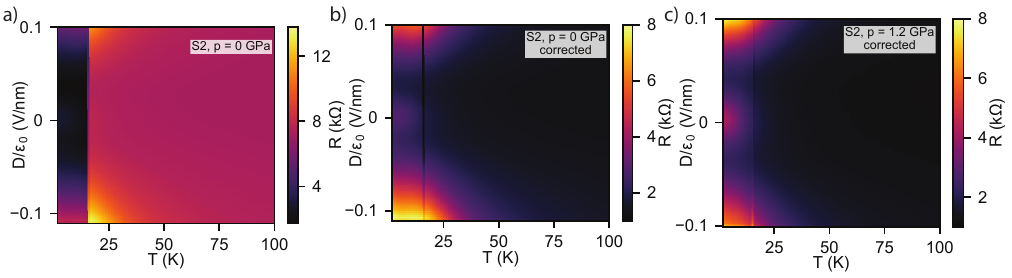}
\caption{\label{fig:activation_data_S2} Temperature dependence of the resistance of device S2 as a function of $D$ at $n=0$ for a) $p=0$. The jump in the resistance around 15 K corresponds to the superconducting phase transition of the NbTiN leads. Corrected data after the contact normal resistance was subtracted for b) $p=0$ and c) $p=1.2$\,GPa.}
\end{figure}

We also performed thermal activation measurements on both devices. For reference, the raw measurement data is shown for dev S2 in Fig.\,\ref{fig:activation_data_S2}.a for $p=0$ obtained while cooling the sample from 100 K to base temperature. Due to the two-terminal geometry of these devices a large jump in resistance is visible that corresponds to the superconducting phase transition of the NbTiN contacts.

In order to extract the band gaps from these measurements additional data processing is required and the normal resistance of the contacts have to be subtracted from the high temperature parts of the data. We subtract $R_{N,S2}=5.90$\,k$\Omega$ and $R_{N,S3}=5.38$\,k$\Omega$ for dev S2 and S3, respectively. The thermal activation data after the subtraction of the contact normal resistance is shown in Fig.\,\ref{fig:activation_data_S2}.b and \ref{fig:activation_data_S2}.c for dev S2 at $p=0$ and $p=1.2$\,GPa, respectively. After the subtraction, a vertical line is visible that originates from the phase transition. We exclude these outliers from further analysis. Although this makes our data analysis less reliable, we can nevertheless extract the SOI strength using the same method as described in section\,\ref{sec:SOI_strength}. This yields $\lambda_t=-\lambda_b=1.6\pm0.4$\,meV for $p=0$ and $\lambda_t=-\lambda_b=2.4\pm0.5$\,meV for $p=1.2$\,GPa and $\lambda_t=-\lambda_b=1.0\pm0.3$\,meV for $p=0$ and $\lambda_t=-\lambda_b=2.0\pm0.4$\,meV for $p=1.2$\,GPa for dev S2 and S3, repsectively. These results also reproduce our findings for dev S1 as a significant increase in the extracted SOI strength is seen in all cases.

\begin{figure}
\includegraphics[width=0.75\columnwidth]{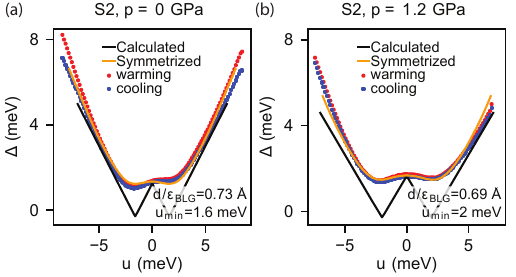}
\caption{\label{fig:activation_gap_S2} Band gaps determined from thermal activation measurements performed while warming up (red) and cooling down (blue) device S2 for a) $p=0$ and b) $p=1.2$\,GPa, respectively. Symmetrized curve with respect to $u=0$ is shown in orange and the band gap calculated from the theoretical model is shown with the solid black line.}
\end{figure}

\begin{figure}
\includegraphics[width=0.75\columnwidth]{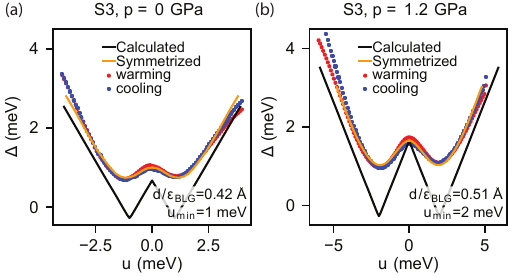}
\caption{Band gaps determined from thermal activation measurements performed while warming up (red) and cooling down (blue) device S3 for a) $p=0$ and b) $p=1.2$\,GPa, respectively. Symmetrized curve with respect to $u=0$ is shown in orange and the band gap calculated from the theoretical model is shown with the solid black line.}
\end{figure}
\newpage

\section{Additional data from Sample S4}
\begin{figure}
\includegraphics[width=0.5\columnwidth]{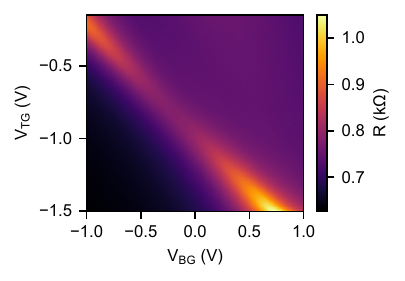}
\caption{\label{fig:gatemaps_noIP} Differential resistance of a two-terminal device as a function of $V_{BG}$ and $V_{TG}$ that shows no signatures of band inversion.}
\end{figure}

As mentioned in section\,\ref{sec:device}, not all samples showed signatures of band inversion. We attribute this to the lack of control over the rotation of WSe$_2$ layers. Based on theoretical predictions\cite{Peterfalvi2022,Li2019}, we argue that for certain rotation angles of the two WSe$_2$ layers (e.g. 0$^{\circ}$), the sign of $\lambda_I^{b}$ and $\lambda_I^{t}$ can be the same which leads to the situation discussed section \ref{sec:theory} and shown in Fig.\,\ref{supfig:sameSOC} where no band gap is present at $D=0$. For reference, in Fig.\,\ref{fig:gatemaps_noIP}, we show measurement data of a two-terminal device measured at 4 K, where no signatures of band inversion can be observed in the resistance map.

\bibliography{inverted_gap_supporting}